\documentclass[10pt,journal,compsoc]{IEEEtran}
\usepackage{cite}
\usepackage{graphicx}
\usepackage{epstopdf}
\usepackage{subcaption}
\usepackage{setspace}
\usepackage[hyphens]{url}
\usepackage{hyperref}
\hypersetup{breaklinks=true}
\usepackage{fancyhdr}
\usepackage{lastpage}
\usepackage[font={small,it},belowskip=-10pt,aboveskip=5pt]{caption}
\usepackage{multicol}
\usepackage{amsmath}
\usepackage{algorithm}
\usepackage{algpseudocode}
\usepackage{balance}
\usepackage{booktabs}
\usepackage{multirow}
\usepackage{siunitx}
\usepackage{xcolor}

\begin{document}
	
\newcommand\comment[1]{{\sffamily [\textbf{comment: #1}]\\}}

\newcommand\Tstrut{\rule{0pt}{2.6ex}}         
\newcommand\Bstrut{\rule[-0.9ex]{0pt}{0pt}}   
\newenvironment{tightcenter}{%
	\setlength\topsep{0pt}
	\setlength\parskip{0pt}
	\begin{center}
	}{%
\end{center}
}

\title{All Infections are Not Created Equal: Time-Sensitive Prediction of Malware \\Generated Network Attacks}

\author
{
	\IEEEauthorblockN
	{
		Zainab Abaid\IEEEauthorrefmark{1}\IEEEauthorrefmark{2},
		Dilip Sarkar\IEEEauthorrefmark{3},
		Mohamed Ali Kaafar\IEEEauthorrefmark{2}\IEEEauthorrefmark{4} and 
		Sanjay Jha\IEEEauthorrefmark{1} 
	}

	\IEEEauthorblockA
	{
		\IEEEauthorrefmark{1}School of Computer Science and Engineering,
		University of New South Wales, Australia\\ 
		\{zainaba,sanjay\}@cse.unsw.edu.au
	}
	
	\IEEEauthorblockA
	{
		\IEEEauthorrefmark{2}CSIRO Data61, Australia.
	}
	\IEEEauthorblockA
	{
		\IEEEauthorrefmark{4}Computing Department, Macquarie University\\
		dali.kaafar@data61.csiro.au\\
	}
	\IEEEauthorblockA
	{
		\IEEEauthorrefmark{3}Computer Science Department
		University of Miami, USA\\
		sarkar@miami.edu\\
	}
}


\IEEEtitleabstractindextext{%
	\begin{abstract}
		Many techniques have been proposed for quickly detecting and containing malware-generated network attacks such as large-scale denial of service attacks; unfortunately, much damage is already done within the first few minutes of an attack, before it is identified and contained. There is a need for an early warning system that can predict attacks before they actually manifest, so that upcoming attacks can be prevented altogether by blocking the hosts that are likely to engage in attacks. However, blocking responses may disrupt legitimate processes on blocked hosts; in order to minimise user inconvenience, it is important to also foretell the time when the predicted attacks will occur, so that only the most urgent threats result in auto-blocking responses, while less urgent ones are first manually investigated. To this end, we identify a typical infection sequence followed by modern malware; modelling this sequence as a Markov chain and training it on real malicious traffic, we are able to identify behaviour most likely to lead to attacks and predict 98\% of real-world spamming and port-scanning attacks before they occur. Moreover,
		using a Semi-Markov chain model, we are able to foretell the time of upcoming attacks, a novel capability
		that allows accurately predicting the times of 97\% of real-world malware attacks. Our work represents an important and timely step towards enabling flexible threat response models that minimise disruption to legitimate users.
		\end{abstract}
	
	}

\maketitle

\IEEEdisplaynontitleabstractindextext

\IEEEraisesectionheading{\section{Background and Motivation}\label{sec:introduction}}

 \IEEEPARstart{M}odern malware exploits the growing number of internet-connected devices and usually takes
the form of botnets,
large networks of infected machines jointly controlled
 by cyber criminals towards malicious ends. Frequently, machines compromised by botnet malware
 have been used to launch various very large scale attacks
 on the internet. In recent years, increasingly large scale network attacks, such
 as 
 distributed denial of service (DDoS) attacks\cite{mirai2016} and point-of-sale credit card breaches\cite{kmartbreach} have been orchestrated using botnets. The potent danger of these 
 attacks lies in the fact that 
 they can cause significant losses immediately after occurring, for example financial
 loss to website owners subjected to DDoS attacks and to users whose private financial information is leaked. Since the dawn of the botnet phenomenon and over the last two
 decades, many techniques have been proposed to detect and mitigate infections and
 contain attacks. Unfortunately however,  much of the damage of an attack is done by the 
 time it is detected and mitigated, even during the first few minutes. 
 Thus, detecting an attack after it is launched
  is no longer sufficient given the scale and potential of attacks. In fact, there is a need for an early-warning system that can
  predict attacks \textit{before they occur}, so that network administrators can take 
  steps to prevent the predicted attacks from materialising. For example,
  hosts that are suspected to engage in attacks in the near future can be monitored,
  manually inspected or taken off the network
  altogether. While the safest response is to auto-block hosts that are likely to engage in 
  attacks, this can lead to user inconvenience when legitimate processes are unnecessarily
  stopped. We argue that it is important to realise that infections do not all behave the same -- attacks may occur at different times within different infections, and the urgency
of the response should depend on the time available to respond to an attack. 
Thus there is also a need for a time-prediction approach along with predicting the occurrence of attacks, so that the response to a prediction can be time-sensitive, and less urgent threats
can first be manually investigated rather than blocking the infected host.
  We emphasise that we are dealing with prediction of outgoing attacks from infected hosts, such as sending of spam emails or sending out malware binaries, and not incoming attacks that cause a previously benign host to become infected with malware, for example a web exploit that delivers malware binary to a previously clean host. 
  
  Publicly available intrusion detection systems (IDS) such as Bro~\cite{paxson1999bro}, Snort~\cite{roesch1999snort} and Suricata~\cite{suricata}, as well as a large volume
  of solutions from prior research, can detect the presence of botnets in network traffic
  with high accuracy, and detect botnet-generated attacks once they have been launched.
In this work, we empirically evaluate whether this existing technology can be harnessed to go a step further and predict and estimate the time of 
upcoming botnet attacks; to the best of our knowledge, no prior research has addressed
this question. Instead, existing work in this domain falls into three broad categories: 
first, solutions for identifying botnet infections or attack behaviour without attempting 
to predict the occurrence of this behaviour; second, prediction of a specific botnet-generated
attack; and third, prediction of general intrusions without a focus on botnets.
  Solutions in the first category 
  are  able to identify  a wide range of botnet infections and a broad spectrum of botnet-generated
  attacks, but cannot \textit{predict} any malicious
  behaviour, including attacks, that may follow in future. 
  Those in the second category are
  limited to predicting
  some specific attacks for which they are designed, 
  such as spam or DDoS; they are not able to generally predict
  any attack from a botnet. Conversely, solutions from the third category of
  intrusion prediction tend to be too general for our purpose (rather than botnet-focused),
  and their potential to predict botnet-generated attacks
  has not been evaluated.  
  Thus, none of these
  solutions can enable botnet-focused prediction of attacks, and 
  consequently cannot apprehend such attacks before they occur. This
  limitation has motivated us to design an \textit{early warning system }
  that (a) works over existing, publicly available IDS tools, (b) can predict any botnet behaviour of interest, including attacks, (c) is not 
  limited to specific attack or botnet types, and most importantly, (d) is sensitive to
  the temporal context within which attacks occur, and can estimate
   the time of occurrence of 
  a predicted attack. We evaluate our solution empirically on a dataset of 
   real botnet-generated traffic to address the question of whether existing IDS technology
   can indeed be successfully harnessed towards this goal of botnet-focused attack prediction.
  
Our approach is based on the intuition that botnet-infected hosts show observable
behaviour other than just attacks. 
Early botnet studies~\cite{abu2006multifaceted} have identified a typical infection sequence (i.e. an ordered
set of behavioural stages) followed by botnets, and traditional detection approaches~\cite{gu2007bothunter}
have effectively detected infections based on the occurrence of this infection sequence. 
In this work, we similarly monitor for the infection sequence, but move beyond traditional
infection detection and instead focus on predicting future attacks 
based on the current behavioural context of a potentially infected host. Specifically, as
infections usually start with incoming exploits (e.g. drive-by download
 or a remote exploit of vulnerable services) resulting in bot malware download, 
  then cause communication with a command-and-control (C\&C) server,
 and finally manifest as outgoing attacks, it is clear that attacks
  occur within a context of distinct behavioural stages. Thus,
 it is possible to fit a model to this context; training the model on real botnet traffic 
 should allow identifying which behavioural stages commonly precede
 attacks and thus allow generating \textit{early attack warnings} when those stages are observed.
 Moreover, capturing the typical 
 temporal distribution of these different stages can allow predicting
 the times of occurrence of attacks given the observation history of an infected host.

 We implement our approach by identifying the botnet infection sequence
 using earlier studies and an analysis of more recent, real-world botnets, 
 and fitting various Markov models to this sequence. 
 We train the models on two datasets comprising traffic from
 a variety of malware families, and propose a methodology to predict future behaviour of an infection
 based on past actions of a host. With a simple Markov chain model, 
 we are able to predict
 attacks with 98\% accuracy. Moreover,
 we show that our methodology can be used
 to predict other behaviour of interest, such as an infected host's
 communication with its C\&C server. Secondly,
 we address the issue of how much history to incorporate into a future
 prediction. We identify that higher-order Markov chains allow incorporating a variable
 amount of history into a prediction decision,
 and empirically investigate whether more history is useful for 
 attack 
 prediction
 by designing a set of higher-order Markov chains and 
 comparing their prediction accuracy with the first-order Markov chain. 
 Furthermore, we build a semi-Markov chain (SMC) to enable a flexible and time-sensitive response to impending attacks, where
 less urgent threats can be handled differently from urgent threats.
 The SMC is able to learn a 
 time-probability distribution from malware traffic, and in combination with the underlying
 Markov chain, is able to generate accurate time-predictions for attacks as well as any other
 future behaviour of botnet-infected hosts. With this model, we can make time 
 predictions regarding when a particular behaviour will occur, with
 accuracy ranging from
 94\% to 100\% for different kinds of behaviour. 
 To the best of our knowledge, this time
 prediction is a novel
 capability not demonstrated in any existing malware detection research.
 In practice, we envision our solution to be deployed as part of a flexible,
 time-sensitive threat response system which can selectively block hosts posing 
 the most urgent threats and allow scheduling manual inspections of hosts
 posing less urgent risks. 
 This would minimise
 user inconvenience resulting from unnecessary disruption of legitimate processes when 
 hosts are blocked simply for being infected.
 
 Overall, this paper makes the following contributions:
 \begin{itemize}
\item An empirical validation of the botnet
attack prediction potential of our proposed approach
 using real traffic from infected machines;
 	\item  An early warning system that can make predictions of future behaviour, including attacks,
 	of botnet-infected hosts with 98\% accuracy, and without the need for prior knowledge of infection.
 	\item A model for the time distribution of botnet events with the novel capability of predicting times of occurrence of future malicious events including attacks with up to 100\% accuracy; 
 	 \item An investigation of whether considering
 	greater historical context is advantageous for the botnet attack prediction problem.
 \end{itemize}

The remainder of this paper is organised as follows. Section~\ref{Sec:related} discusses related work in this domain. Section~\ref{Sec:Theory} defines some key terminology. In Section~\ref{sec:model} we design a model to capture the
botnet infection sequence, and in Section~\ref{sec:training}, we instantiate different attack
prediction models. 
Section~\ref{sec:selfTransitions} presents some empirical analysis of the data relevant to our prediction
approach, and Section~\ref{Sec:prediction} outlines our experimental methodology and evaluation results.
Section~\ref{sec:feasibility} discusses the practical feasibility of our approach, and
Section~\ref{sec:concl}  concludes the paper.

\section{Related Work}
\label{Sec:related}
While identification of botnet infections and attacks has been a
well-researched problem, a solution for \textit{predicting} diverse 
botnet attacks while
remaining independent of botnet and attack type has 
remained elusive. In addressing this gap, we have drawn 
inspiration from three broad categories of prior research: (a)
identifying diverse
botnet infections but without a focus on prediction, (b)
predicting 
specific botnet attacks (i.e.solutions that are not generic to diverse attack types),
and (c) intrusion prediction without a focus on botnets. We now discuss some key approaches
belonging to each of these categories and also review the use of Markov models similar to ours
in prior intrusion
detection research, and compare our work with this existing research.

In the first category of research, the idea of a  bot ``lifecycle", first proposed in~\cite{abu2006multifaceted}, has been used to identify botnet infections, by
monitoring for the various behavioural stages that a botnet
infection exhibits; infections are declared if multiple such stages are observed on a 
host in a particular sequence~\cite{gu2007bothunter,gu2008botminer,haq2015sdn}. 
An example of this approach is found in BotHunter~\cite{gu2007bothunter}, where
 a host is classified as infected if it engages in communication sessions
belonging to multiple ``lifecycle" stages, such as malware downloading, 
C\&C communication, and performing attacks. This approach forms
an inspiration for our work as we too monitor hosts for signs of various bot lifecycle stages. However, instead of stopping at declaring that an infection
has occurred, we go a step further and use the identified behaviour to predict what
the future behaviour of the host is likely to be and whether (and when) it will engage
in attacks. 

In the second category of work on predicting specific attacks, only a limited
amount of prior research exists. One scheme~\cite{hirayama2015target} uses the increase in traceroute packets in the network prior to a target link DDoS attack
to predict the attack itself, detecting the preparation stage before the attack occurs. 
This approach however is restricted to a specific attack.
To the best of our knowledge, no current research has investigated a botnet-focused attack prediction solution not restricted to specific attack types.


Closest to our work is the third category of research on intrusion alert prediction, i.e. training a model on an observed sequence of alerts from an IDS and using it to predict future alerts~\cite{thanthrige2016intrusion, fava2008projecting, holgado2017real}. \cite{thanthrige2016intrusion} clusters intrusions alerts generated by the Snort IDS in training traffic, and trains an HMM where the alert clusters are used as the observable symbols emitted by the hidden states, and calculates the most likely next alert category to be generated. While this category of work matches our goal of predicting future malicious activity, our work differs in its focus on botnet-generated attacks as opposed to general intrusion alerts. Our approach first maps alerts to known stages typical of botnet behaviour, discarding all alerts that do not signify botnet activity, which greatly reduces the amount of data to be processed by the system in both the training and deployment phase. To the best of our knowledge, ours is the first work to empirically evaluate the effectiveness of intrusion prediction in predicting botnet attacks.

Finally, our current work draws inspiration from a large body of prior research that has proposed various Markov models in the context of intrusion detection. 
These include single-order as well as higher-order 
Markov chains, semi-Markov chains (SMCs), and Hidden Markov Models (HMMs).
A typical example of single-order Markov chains in intrusion detection is~\cite{ye2000markov}, which trains a Markov chain
using events generated by an audit system on a host. In the deployment phase, event streams 
generated by hosts that are dissimilar to the learned benign model
are classified as attacks.  
Similar approaches are presented in \cite{ye2001probabilistic, estevez2005detection}. 
Recently, Markov chains have been applied to Android malware detection, where
the sequence of API calls~\cite{mariconti2016mamadroid} or system service calls~\cite{salehi2017android} in an application are used to build a Markov chain, which is then used as the feature vector for that application. A classification algorithm (e.g. Random Forest) is trained over  Markov chains of malicious and benign applications in a training set and 
used to classify future applications as malicious or benign. 
This approach is inherently different
from ours as Markov chains are used as feature vectors for classifying applications rather than
to predict the next event in a sequence, and hence cannot be used for attack
prediction. 
HMMs have been used often for anomaly detection, where the anomalous
behaviour to be detected
is represented as a sequence of symbols, and the states of the model that are not otherwise observable,
are assumed to emit those symbols. The observed data is then checked for similarity to the trained
model to detect anomalies. This approach has been applied to botnet detection~\cite{gobeldetecting, venkatesh2013http, msthesis}, and recently also to program anomaly detection~\cite{xu2016sharper}.
 However, these approaches are restricted to
identifying infections or malicious behaviour that has already started, rather than
predicting
the future behaviour of potentially infected hosts, which is the focus of our work.

One issue that arises when
using Markov models for prediction of future behaviour is the amount of history
to consider when making a prediction.
Higher-order Markov chains~\cite{raftery1985model}
in which the last \textit{m} states determine the next state, incorporate more history than single-order chains, and
have been used
in predicting web browsing behaviour of users given their past observed history of web requests~\cite{borges2007evaluating, deshpande2004selective}. Intrusion detection applications are less common;
one such work~\cite{ju2001hybrid} partially
uses a higher-order Markov chain, along with other models,
to differentiate between command sequences from legitimate system users and intruders.
Another issue in prediction is that one may be interested not just in the next transition but also in its time of 
occurrence.
SMCs~\cite{barbu2008semi} address this by incorporating a time probability
distribution in the Markov chain. While SMCs have not been used in \textit{predicting}
intrusions, an inspiration for our research is~\cite{xie2009large} which
uses a hidden SMC to differentiate between legitimate and anomalous web-browsing behaviour (that could indicate DoS attacks) based on the requests received by web servers. 
To the best of our knowledge, neither
higher-order Markov chains nor SMCs have not been used in botnet
attack prediction; part of our work contributes to this gap by systematically performing such an investigation.

In summary, we believe that there is no existing approach, whether using Markov chains or otherwise, that evaluates whether botnet attacks, as well as their times of occurrence,
can be accurately predicted while remaining independent of botnet or attack type. 

\section{Single-Order, Higher-Order and Semi-Markov Chains}
\label{Sec:Theory}

Markov chains are used to model systems that randomly move among a set of states. In general,
a Markov chain is represented as a state transition diagram with a set of states $[S]$, with the paths between the states weighted by the probability of moving from one state to another; each transition from a state $S_i$ to a state $S_j$ has a probability $t_{i,j}$ of occurrence. We now discuss some types of Markov chain
that we use for developing our attack prediction models.

\subsection{Single-Order Markov Chain}
\label{subsec:singleMarkov}
A single-order Markov chain is defined by a strict memoryless property, i.e. the next state depends only
on the current state; neither older states nor any other variables, such as the time spent in the current
state, have any effect on determining the next state.
We briefly define some terminology that we later refer to in developing our single-order Markov chain model. An \textit{irreducible} Markov chain is one in which all states can communicate -- i.e. for each pair of states $S_i$ and $S_j$, there exists a path from  $S_i$ to $S_j$ and vice versa. 
A state is \textit{periodic} if it can only be returned to at time values that are multiples
of an integer greater than 1, e.g. at time $t$ $=$ $3,6,9$ and so on. The period of a state
is calculated by taking the greatest
common denominator (GCD) of all possible times of return to the state.
States with a period of $1$, such as those with a self-transition, are \textit{aperiodic}, and a chain for which all states are aperiodic is an aperiodic chain. For an irreducible Markov chain, it is possible to define a \textit{stationary distribution} representing the long-term probabilities of being in each of the states, i.e. $P(S_i) $ for all states $S_i$ as time grows large. This distribution should satisfy the
 properties $P_{j} = \sum_{i \in [S]}P_{i}t_{i,j}$
	and	$\sum_{i\in [S]}P_{i} = 1$,
	where 
$[S]$ is the set of states in the model,
$P_{i}$ is the stationary probability of being in state $S_i$, and
$t_{i,j}$ is the probability of going from state $S_i$ to state $S_j$.
 If the chain is aperiodic, then this stationary distribution is also the \textit{limiting distribution} for the Markov chain.
Finally, a \textit{reversible} Markov chain has to satisfy for all pairs of states $S_{i}$ and $S_{j}$ the property 
$P_{i}t_{i,j} = P_{j}t_{j,i} \forall i,j$ 	
where $P_{i}$ and $t_{i,j}$ are as defined above.

\subsection{Higher-Order Markov Chains}
In a single-order Markov chain, the probability of the next state depends only on the current 
state; an $m^{th}$ order chain extends this dependence, such that the probability of the next state 
depends on the last $m$ states up to and including the current
state\cite{raftery1985model}.
Higher order Markov chains are useful in situations where the prediction of the next event requires more
history than just the previous event, i.e. where different combinations of multiple events carry different
influences on the future. Therefore, higher-order chains appear to be a good fit to the botnet attack
prediction problem. Intuitively, it seems possible 
that different combinations and sequences of botnet events may 
lead to attacks with different probabilities. For example, a certain botnet may behave as follows: when a host downloads a malicious binary, it first connects to the C\&C server, and then downloads further 
executables containing attack instructions. Then it engages in an attack. Thus, attacks should be
predicted when the sequence ``$Binary \: Download \rightarrow C\&C \thinspace Communcation \rightarrow Binary\thinspace Download$ " is observed, rather than
after only the Binary Download event is observed.
However, this distinction cannot be captured with a first-order Markov chain. 
Applying a higher-order Markov chain to the 
attack prediction problem would allow us to investigate whether an improvement in attack prediction
can be achieved by looking farther back into the past when making a prediction.


\subsection{Semi-Markov Chains}
Regardless of its order, a Markov chain ignores any effect of the time spent in the current state on
determining the next state, which we refer to as the \textit{state holding time distribution}. 
However, in the context of attack prediction,
it is important to not just detect when the Attack state will occur, but also to estimate how much
time will be spent in the current state before it does occur. The response to an attack warning can then be 
varied according to the amount of time available. 
For example, if an attack is predicted to occur within a 
minute of the warning, the best response may be to quarantine the potential attacker host immediately. 
However, if the attack is predicted to occur after an hour, manual verification and response 
to the problem may be possible. The advantage of predicting the time of an attack would be to  reduce the number of auto-quarantine responses to warnings, which are not desirable as completely denying network connectivity to 
infected hosts would pose a serious usability problem to legitimate users. 
Legitimate processes should be allowed to run while
attacks are investigated and blocked. A semi-Markov chain~\cite{barbu2008semi} offers a solution to estimate the holding times of states. 

A semi-Markov chain can be applied over a standard Markov chain. The 
holding times for the states of the \textit{embedded} Markov
chain are assumed to be random variables obeying a distribution,
which
can be estimated from training data. Thus, a semi-Markov chain is defined primarily 
by an underlying Markov chain with a transition
probability matrix $T_{ij}$, and a time probability distribution $F_{ij}(t)$ which describes the probability
of each possible transition $t_{ij}$ occurring within a varying time interval $(0,t]$. 
In a continuous-time semi-Markov chain, the time $t$ is continuous and may be any real number, while a discrete time Semi-Markov chain divides the time into specific intervals where transitions occurring within a particular interval $(t_x, t_y]$ are all grouped together. In this work, we design a discrete-time semi-Markov chain,
where it is possible to determine, for a given interval $I_n$,  the probability
of a transition from any state $S_i$ to any state $S_j$ within the interval $I_n$. Thus, when we predict
an attack, we can also predict its approximate time of occurrence.

\section{Designing a Model to Capture Malware Infection Sequences}
\label{sec:model}

In this section, we present a model for capturing the typical sequences observed within
malware infections. 
Defining such a model is a pre-requisite for the core 
contribution of our work, i.e. predicting
 attacks, as it represents the underlying
state diagram of all three Markov
chain models that we use for prediction.


\subsection{Proposed Model}
\label{subsec:model}
\begin{figure}
	
	\centering
	\includegraphics[width=0.8\columnwidth]{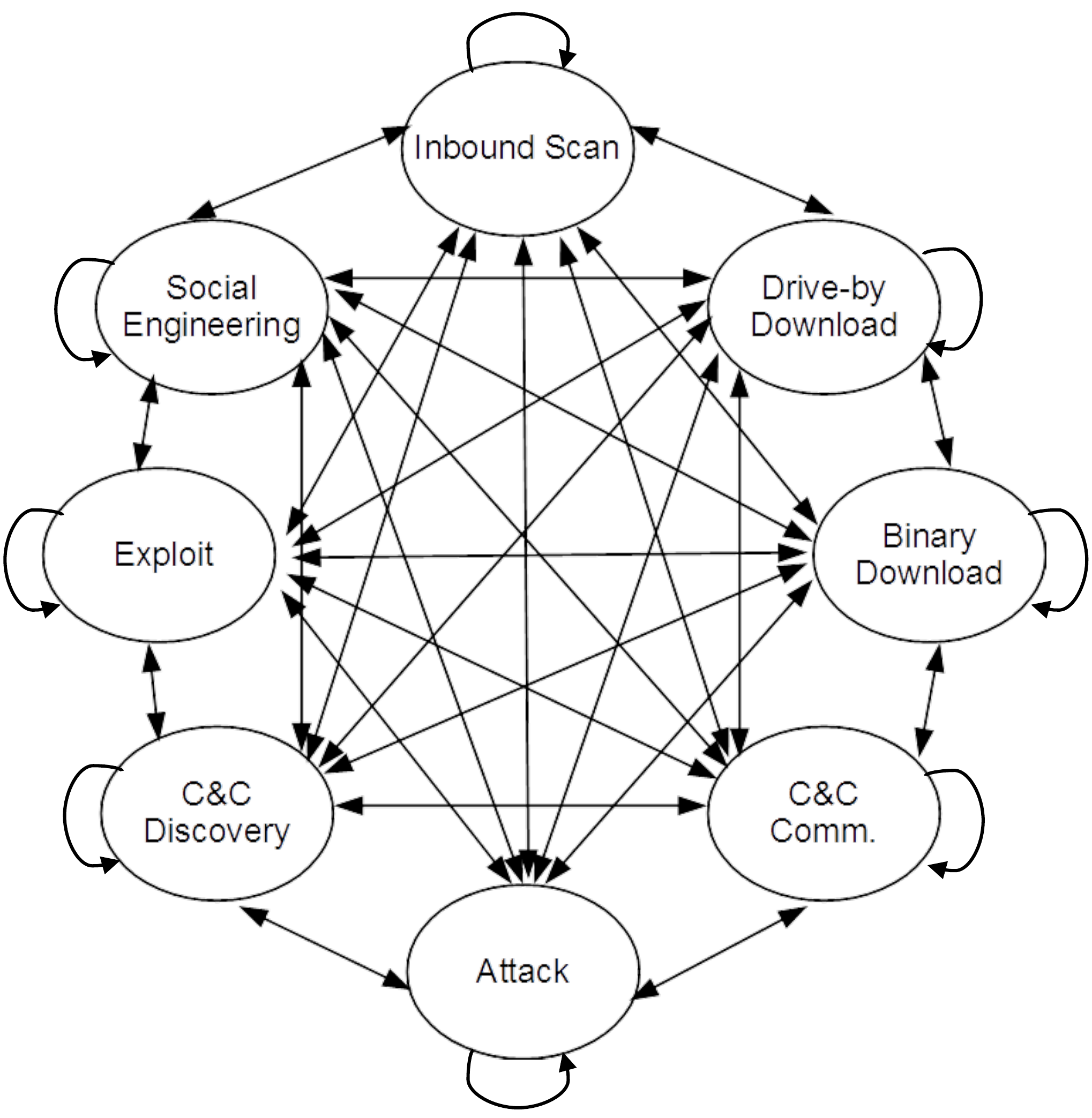}
	\caption{State transition diagram modelling botnet infection stages.}
	\label{fig:stateDiagra}
\end{figure}

Prior studies\cite{abu2006multifaceted,gu2007bothunter,silva2013botnets} have suggested that bots follow a well-defined sequence, beginning with a social engineering attack or an exploit, which results in a malicious binary download, followed by communication with a C\&C server and then launch of an outgoing attack, such as spam generation or DDoS attacks. We examine the behavioural stages exhibited by a range of modern
botnets, including recent mobile botnets, to verify that this sequence remains invariant, and
conclude that outgoing attacks occur within a context that can be modelled. We capture this context as a state transition diagram, shown in Fig.~\ref{fig:stateDiagra}, with each state representing a different stage of infection.

\begin{table*}
	\centering 
	\scriptsize
	\vspace{1em}
	\caption{Description of states of the Markov Chain and mapping \texttt{Dridex 120} behaviour to each state (where applicable).}
	
	\begin{tabular}{ |l || p{0.375\textwidth} || p{0.375\textwidth} |} 
		\hline
		\textbf{State} & \textbf{Description of Behaviour} & \textbf{Example of Dridex Behaviour} \Tstrut\\ 
		\hline
		\hline 
		
		Inbound Scan & An incoming port scan on any port of the host, characterised by the same port being scanned on many hosts (horizontal scan) or many ports being scanned on one host (vertical scan).
		& N/A \Tstrut
		\\
		\hline
		
		\Tstrut Social Engineering &  Human users are tricked into compromising
		on usual security practices, e.g. clicking on an intriguing (but malicious) spam link in an email message. & Victim receives email with a malicious Microsoft Word or Excel attachment.\\
		\hline
		
		\Tstrut Drive-by Download & Malicious executables are downloaded without the user's consent during a web browsing session. & User opens the email attachment, triggering execution of embedded macro, which then causes download of an intermediate dropper. \\
		\hline
		
		\Tstrut Exploit & Any attack targeting application or OS vulnerabilities for gaining ``backdoor'' access or remote execution privileges
		on the victim machine, or web browser. & N/A\\
		\hline
		
		\Tstrut Binary Download &  The actual download of a malicious piece of code or an executable, which may be disguised as 
		legitimate software or an image or document. &  Dropper executes, and downloads and runs the actual bot binary.\\
		\hline
		
		\Tstrut C\&C Discovery & When the downloaded binary is run and attempts to contact the botnet's C\&C server. Discovery behaviour may include contacting a list of servers one by one over HTTP, opening
		multiple P2P connections, or a large number of DNS queries for a list of possible domains. & Sends an HTTP Post request to a specific hard-coded IP address.\\
		\hline
		
		\Tstrut C\&C Communication & Exchanges with the discovered its C\&C server, over IRC, HTTP, or in some cases, custom protocols. & If HTTP POST is successful, bot receives configuration and instructions regarding which websites to target for redirection attacks.\\
		\hline
		
		\Tstrut Outgoing Attack & Outgoing traffic that can damage other entities; e.g. port scanning, information stealing, phishing,
		DoS or
		spam. & When victim visits a banking website included in the configuration file, the bot can perform redirection to a phishing website and credential stealing. \\
		
		\hline
	\end{tabular}
	\label{Tab:stages}
\end{table*}

\emph{States of the Model:}
Table~\ref{Tab:stages} describes the behaviour each state in our model represents and exemplifies each stage with a recent banking trojan called
Dridex, mapped to the relevant states of our model. We derive this information from an analysis of \texttt{Dridex 120} published by CISCO~\cite{ciscoblog}. Our model is one
of many possible representations of botnet behaviour; for example,
each of its states can be broken down into multiple states by considering
different variations of a behavioural category. However, we restrict our model to eight states that condense all variations of each stage
of behaviour (including attacks) into one high-level state. We believe that this model provides comprehensive coverage of behaviour exhibited by a typical modern botnet infection. 

\emph{Full Mesh Structure of the Model:}
Unlike previous work, in which states are linked in a logical flow that the infection should proceed in, we link the states in a full mesh to avoid imposing our own  assumptions about the infection sequence on the model. The sequence in which the stages occur can and does vary considerably between malware types. In fact, botnet traces used in the experimentation show significant variation in the sequence of stages observed before attacks occur.  The full mesh structure ensures that a compromised host does not have to follow a set sequence of stages for our attack prediction to work, as long as at least some of the stages can be detected on the host, and the training data contains a few different variations in the sequence.

\subsection{Current Model Limitations}
\label{subsec:modelLimitation}

Applying the comprehensive model of Fig.~\ref{fig:stateDiagra} entails being able to detect activities falling within each state of the model. To the best of our knowledge, this is not possible with existing tools. While tools such as Suricata~\cite{suricata}, Bro~\cite{paxson1999bro} and Snort~\cite{roesch1999snort}, are capable of  detecting several such activities, at present there is no rule set or script publicly available with these tools that contains detection logic for all the behavioural stages in our model. Currently, we use Snort as our detection tool, with botnet detection rule sets from Emerging Threats~\cite{emerging} as well as from BotHunter~\cite{gu2007bothunter}, an industry-standard botnet detection IDS, a combination that allows us to detect a greater variety of malicious activities than any other existing tool. 
We acknowledge that our model is IDS-dependent in its effectiveness. If the underlying IDS misses any of the malicious behaviour that a host engages in, it would compromise the attack prediction accuracy of our model. However, improving the IDS is beyond the scope of our work, which simply intends to demonstrate that applying our models over any existing IDS can allow
extending its detection capability to prediction.
We now merge the Drive-by Download
state into the Exploit state, and the C\&C Discovery state into the C\&C Communication state, as Snort does not always distinguish between these pairs.
We remove  the social engineering state as
detecting it entails being able to link a social media message or an email to a downloaded piece of malware,
which is not possible with existing tools.
We remove the Inbound Scan state as it is never detected within our current dataset. 
For the remainder of this paper, we work with a four-state model comprising the states \textit{Exploit, Binary Download, C\&C Communication, Attack} linked in a full mesh. However, all our methodology is applicable as it is to the full model, including the calculations and the proofs of validity that follow in the next section.

\section{Training Attack Prediction Models}
\label{sec:training}
We now describe how we instantiated and trained three attack prediction models -- a single-order
Markov chain, a set of higher-order Markov chains, and a semi-Markov chain -- on two  malware 
datasets.

\subsection{Datasets}
\label{subsec:data} 

We used two traffic traces containing known malware activity to instantiate our model. Note that it is only for evaluation purposes that we use datasets of known infected hosts; in a practical deployment there is no need to pre-label the hosts as infected or clean.

\begin{table*}
	\centering
	\vspace{1em}
	\small
	\caption{Example of sequences of behavioural stages observed in dataset.}
	\begin{tabular}{ | p{0.10\textwidth} || p{0.85\textwidth} |}
		\hline
		\Tstrut \textbf{IP} & \textbf{Event Sequence} \\ \hline
		\Tstrut a.a.a.a & 2:CNC, 3:CNC, 4:ATTACK, 7:EXPLOIT, 9:CNC, 13:BINARY, 16:EXPLOIT, 18:ATTACK, 23:CNC\textbf{...} \\ \hline 
		\Tstrut b.b.b.b	& 0:EXPLOIT, 6:EXPLOIT, 7:BINARY, 19:ATTACK, 23:CNC, 26: BINARY, 27:ATTACK, 29:EXPLOIT\textbf{...} \\ \hline
		\Tstrut c.c.c.c	& 1:ATTACK, 2:ATTACK, 6:EXPLOIT, 7:BINARY, 9:CNC, 10:ATTACK, 11:CNC, 14:EXPLOIT, 18:BINARY\textbf{...} \\
		\hline
	\end{tabular}
	\label{Tab:snapshot}
\end{table*}

\subsubsection{The SysNet Dataset}
The SysNet trace has been generated by the SysNet lab~\cite{sysnet} in July 2013 and contains traffic from ten infected hosts. It was generated
by running ten bot binaries in separate virtual machines: four variants of Pushdo, two variants of Sality, and one each of Kolabc, Virut, Dorkbot, and Bobax.
This covers HTTP, IRC, and P2P-based bots that engage in a range of attacks including sending spam and outbound scanning.
Full packet traces from the virtual machines were captured for 24 hours using Wireshark on host machines.

\subsubsection{The ISCX  botnet Dataset}
This trace was collected by researchers at the ISCX Laboratory at UNB, Canada and made available publicly. It contains full traces of 30 infected machines, nearly half of which are infected by IRC botnets and the remainder by a variety of P2P and HTTP botnets, including variants of Zeus, Virut, NSIS, Storm, and Zero Access among others. 
Interested readers are referred to \cite{beigi2014towards} for details of the trace. 
Unfortunately, not all hosts showed attack behaviour detectable by current intrusion
detection tools; therefore, we dropped such hosts and selected a total of $15$ hosts' traces
for inclusion in our experimentation, on the basis that they showed evidence of engaging in attacks.
The duration of the selected traces varies from a few minutes to nearly 3 days. Because of the short duration of several of these traces, we found that this dataset by itself is unsuitable for splitting into training and testing portions for evaluating the model. Thus,
we combined traces from both datasets into a single dataset comprising $10$ hosts from the SysNet trace
and $15$ hosts from the ISCX trace. 

For training the Markov chain, the traces were mapped to state sequences comprising the states in our model, 
i.e. Exploit, Binary Download, C\&C Communication and Attack. 
We ran Snort, which runs over traffic and generates alerts when a rule is triggered, over our data using botnet detection rules from Emerging Threats and BotHunter; next
we post-processed the Snort logs to identify alerts belonging to one of the states in our model, discarding irrelevant alerts.
We then generated a timestamped sequence of states observed in each host's network traffic. A snapshot of the data used to instantiate our model is shown in Table~\ref{Tab:snapshot}. The timestamp represents minutes from start of the trace; for example, ``\texttt{2:CNC}'' indicates that a C\&C Communication event was detected at time 2 minutes from the start of the trace. We next describe how we use this data to train our prediction models.


\subsection{Instantiating the Single-Order Markov Chain Model}
\label{subsec:instantate}
The first step in developing a single-order Markov chain  is
to instantiate our proposed model (from Section~\ref{subsec:modelLimitation})
 with the state sequences extracted from the dataset
 and verify that it represents a valid
Markov chain, according to the properties defined in Section~\ref{subsec:singleMarkov}. 
To this end, we first discuss
the irreducibility and aperiodicity of our model, calculate
its transition and stationary probabilities to empirically verify that 
the stationary distribution
represents a valid Markov chain,
and finally verify its reversibility.

\subsubsection{Irreducibility and Aperiodicity}
Showing our Markov chain to be irreducible requires no proof. As discussed in Section~\ref{subsec:singleMarkov}, irreducibility simply refers to a chain where all states can communicate with each other. 
The states in our model are connected in a full mesh so this is indeed the 
case. Also, as all states in our model have self-transitions; 
therefore the Markov chain is aperiodic. 
This allows us to label the stationary distribution as the limiting
distribution for this chain, which can only be valid if the chain is aperiodic.

\subsubsection{Verifying Properties of the Stationary Distribution}

To empirically verify that the stationary probability distribution that we have obtained
 fulfils the conditions for a Markov stationary distribution outlined in Section~\ref{subsec:modelLimitation},
  we need to derive the stationary probabilities representing long-term likelihood of being in each state.
To this end, we use the flow balance condition of a Markov model,
i.e. the probability of leaving a state must equal that of entering it.
This allows us to formulate a balance equation for each state of the model.
In general for each state $S_i$ that has incoming transitions from a set of $n$ states and has
outgoing transitions to a set of $m$ states:

\begin{equation}
\sum\limits_{k=1}^{n} P_{k}t_{k,i} = \sum\limits_{j=1}^{m} P_{i}t_{i,j}
\end{equation}
where 
$P_{i}$ is the limiting probability of being in each state, and
$t_{i,j}$ is the transition probability from state $S_i$ to state $S_j$.
Based on this we define the following flow balance equations for each of the states in the model
from $S_{1}$ to $S_{4}$ respectively:
\vspace{-1mm}
\begin{equation}
P_{2}t_{2,1} + P_3t_{3,1} + P_{4}t_{4,1} = P_{1}t_{1,2} + P_{1}t_{1,3} + P_{1}t_{1,4} 
\end{equation}
\vspace{-2mm}
\begin{equation}
P_{1}t_{1,2} + P_{3}t_{3,2} + P_{4}t_{4,2} = P_{2}t_{2,1} + P_{2}t_{2,3} + P_{2}t_{2,4}
\end{equation}
\vspace{-3mm}
\begin{equation}
P_{1}t_{1,3} + P_{2}t_{2,3} + P_{4}t_{4,3} = P_{3}t_{3,1} + P_{3}t_{3,2} + P_{3}t_{3,4}
\end{equation}
\vspace{-2mm}
\begin{equation}
P_{1}t_{1,4} + P_{2}t_{2,4} + P_{3}t_{3,4} = P_{4}t_{4,1} + P_{4}t_{4,2} + P_{4}t_{4,3}
\label{eq:last}
\end{equation}
where $P_{i} \:\forall \: i_{1..4}$ and $t_{i,j}$ are are defined above.

In order to calculate the value $P_i$ for each state $S_i$, we first need transition probabilities $t_{i,j}$.
Therefore we first generate a transition probability matrix $T$ from the dataset of state sequences 
with each cell $[i,j]$ representing the 
probability of a transition between states $S_{i}$ and $S_{j}$.
A common approach to
calculate the
transition probability matrix 
is to build a training sequence of states by observing the system for some length of time,
and then
generate the probability $t_{i,j}$ for  each pair of states $S_i$ and
$S_j$ as follows:
\begin{equation}
\frac{\sum N_{i,j}}{\sum\limits_{k\in [S]}N_{i,k}}
\end{equation}

where $\sum N_{i,j}$ represents the  number of transitions observed from state $S_i$ to $S_j$, and
$\sum\limits_{k\in [S]}N_{i,k}$ represents the total number of transitions observed from state $S_i$ to any
state in $[S]$.
For our model and dataset, T is a $4\times4$ matrix and contains the following values:\\ 
\begin{center}
	$
	T = 
	\begin{pmatrix}
	0.682 & 0.030 & 0.033 & 0.254 \\
	0.035 & 0.426 & 0.527 & 0.012 \\
	0.0001  & 0.001  & 0.926 & 0.073  \\
	0.001 & 0.00001 & 0.099 &  0.899 
	\end{pmatrix}
	$
\end{center}
\Tstrut 
The states represented by each row and column are Exploit, Binary Download, C\&C Communication and Attack, respectively. To solve for $P_i$, we use the values from the
matrix $T$ and replace Equation~\ref{eq:last}  with the conservation of total probability equation
$\sum_{i \in S}P_{i} = 1$.
This allows us to obtain the stationary distribution for our dataset shown in Table~\ref{Tab:probs}; it appears that C\&C Communication is the activity that infected hosts tend to engage in most 
frequently.

\begin{table}
	\centering
	\small
	\vspace{1.5em}
	\caption{Stationary probability distribution for the dataset showing long-term probability of
		being in each state.}
	\begin{tabular}{ | p{0.44\columnwidth} || p{0.44\columnwidth} |}
		\hline
		\Tstrut \textbf{State} & \textbf{Probability}\\
		\hline\hline
		\Tstrut Exploit &   $ P_{1} = 0.0025$ \\
		Binary Download & $P_{2} = 0.0015$\\
		C\&C Communication &  $ P_{3} = 0.5739$\\
		Attack & $P_{4} = 0.4222$ \\
		\hline
	\end{tabular}
	\label{Tab:probs}
\end{table}

We verify empirically that the stationary probability distribution holds true to both
conditions 
specified in Section~\ref{subsec:singleMarkov}, i.e. the conservation of total probability 
and the condition
$	P_{i} = \sum_{j \in S}P_{j}t_{j,i}$.
The calculations are straightforward and we do not show them here.

\subsubsection{Reversibility}
\label{subsubsec:transMatrix}
Reversibility requires that for all states $S_i,S_j$:
$P_{i}t_{i,j} = P_{j}t_{j,i}$.
Using the calculated stationary probabilities and the matrix $T$, we verify empirically
that this property holds for all states in the model. We omit the calculations for 
brevity. 

\begin{table*}
	\centering
	\small
	\vspace{1em}
	\caption{Time intervals for Semi-Markov Chain, in minutes.}
	\begin{tabular}{ |c || c | c | c | c | c | c | c | c |}
		\hline
		\Tstrut \textbf{ID} & 1 & 2 & 3 & 4 & 5 & 6 & 7 & 8 \\
		\hline\hline
		
		\Tstrut \textbf{Interval} 
		& $ 0 < t \leq 1$ 
		& $ 1 < t \leq 10$ 
		& $ 10 < t \leq 20$
		& $ 20 < t \leq 30$
		& $ 30 < t \leq 40$ 
		& $ 40 < t \leq 50$
		& $ 50 < t \leq 60$
		& $ t > 60 $ \\
		\hline
	\end{tabular}	
	\label{Tab:intervals}
\end{table*}

\subsection{Instantiating the Higher-Order Markov Chain}
\label{subsec:instantateHigh}
We build a set of higher-order Markov chains up to order 9, and train each chain on
our data as follows.
For a given chain of order-$m$, we learn a
transition probability matrix $T^{(m)}$ -- the difference from the order-$1$ matrix shown in 
Section~\ref{subsec:instantate} is that the rows
no longer correspond to single states, but instead to ordered length-$m$ sequences 
of the states in $[S]$. We refer to the sequence represented by row $i$ as $seq_i$.
Each cell $t_{i,j}$ of the matrix is calculated as follows:
\begin{equation}
\frac{\sum N^{(m)}_{seq_i,j}}{\sum\limits_{k\in S}N^{(m)}_{seq_i,k}}
\end{equation}
where
$N^{(m)}_{seq_i,j}$ is the number of transitions from the sequence $seq_i$ to a state $j$, and
$\sum\limits_{k\in S}N^{(m)}_{seq_i,k}$ is the number of transitions from the sequence $seq_i$
to any state in [S].

\subsection{Instantiating the Semi-Markov Chain}
\label{subsec:instantateSemi}

We apply a discrete-time semi-Markov chain over the same underlying Markov chain as defined in Section~\ref{subsec:instantate} and empirically
estimate a holding time distribution, $F_{ij}(t)$ as follows.
As the 
main objective of estimating time is to be able to respond to attacks differently depending on
the length of time available, we divide the possible holding times into eight intervals
representing various lengths of time available for taking preventative measures against 
predicted attacks. A network
administrator receiving a warning that an attack will occur after a particular length of
time can decide on a suitable preventative measure  
according to the available time.
The set of intervals $I$ is shown 
in Table~\ref{Tab:intervals}. 
The first interval $(0,1]$ is just a minute long -- this is 
especially adapted to our dataset where we notice a very high proportion of
states occurring very close together (less than a minute apart). If an attack is 
predicted to occur within this interval, the only possible preventative action can be to
quickly -- and perhaps automatically --  block
all traffic from the attacking host. The next six intervals
are ten minutes each, and the eight interval represents any length of time over an hour. Attacks
predicted to occur within the eighth interval can almost certainly be manually investigated.


Having defined the set of intervals $I$, we represent $F_{ij}(t)$ as
a set of eight transition probability matrices, one pertaining to each 
time interval $I_n$ where $1 \leq n \leq 8$. We then
learn the eight matrices 
using an approach similar to learning the transition probability matrix $T_{ij}$ discussed in Section~\ref{subsec:instantate}. That is, for a
given interval $I_n$, the element $i,j$ of the matrix corresponding to $I_n$ represents the probability
of a transition from state $S_i$ to state $S_j$ within the interval $I_n$, and is calculated as follows:

\begin{equation}
\frac{\sum N_{i,j}(I_n)}{\sum\limits_{k\in [S]}N_{i,k}}
\end{equation}
where 
$\sum N_{i,j}(I_n)$ represents the number of transitions observed from state $S_i$ to state $S_j$
within the interval $I_n$, and
$\sum\limits_{k\in [S]}N_{i,k}$ represents the total number of 
transitions from the state $S_i$ to any other state in any time interval.

After learning the distribution $F_{ij}(t)$, 
the holding time for each state ($H_{S_i}$) can then be estimated in terms of a probability for a given
time value $t_x$ as follows:
\begin{equation}
H_{S_i}(t_x) = P\{T_{S_i} \leq t_x\} = \sum\limits_{j\in [S]} F_{ij}(t_x)P_{ij}
\label{eq:holdTime}
\end{equation}
where
$H_{S_i}$ is the time spent in state $S_i$ before any transition, and
$ F_{ij}(t_x)$ is the probability of the next transition from state $S_i$ to state $S_j$ occurring within time $t_x$
and can be obtained from the learned $F_{ij}(t)$ distribution after mapping $t_x$ to one of the pre-defined intervals in $[I]$;
$P_{ij}$ is the  embedded transition probability from state $S_i$ to $S_j$.

\section{An empirical analysis of self-transitions}
\label{sec:selfTransitions}

\begin{table}
	\centering
	\small
	\vspace{0.9em}
	
	\caption{Statistical properties of \textbf{duration} (in minutes) of self-transitions for each state.}
	\begin{tabular}{ |p{0.3\columnwidth} || p{0.08\columnwidth}| p{0.08\columnwidth} | p{0.08\columnwidth} | p{0.08\columnwidth}| p{0.08\columnwidth}|}
		\hline
		\Tstrut \textbf{State} & \textbf{Min.} & \textbf{Max.} &   \textbf{Mode} & \textbf{Mean}& \textbf{Stdev.} \\
		\hline\hline
		\Tstrut Exploit & $<1$ & $6$ & $<1$ & $1.3$ & $2.2$\\
		Binary Download & $<1$ & $14$ & $<1$ & $0.75$ & $3.12$\\
		C\&C Comm. & $<1$  & $1437$ &  $<1$ & $2.7$ & $47.1$\\
		Attack & $<1$ & $1420$ & $<1$  & $5.6$ & $63.2$ \\
		\hline
	\end{tabular}
	\vspace{1mm}
	\label{Tab:durationStats}
\end{table}

\begin{table}
	\centering
	\small
	\vspace{1em}
	\caption{Statistical properties of the \textbf{number} of self-transitions for each state.}
	\begin{tabular}{ |p{0.3\columnwidth} || p{0.08\columnwidth}| p{0.08\columnwidth} | p{0.08\columnwidth} | p{0.08\columnwidth}| p{0.08\columnwidth}|}
		\hline
		\Tstrut \textbf{State} & \textbf{Min} & \textbf{Max} & \textbf{Mode} & \textbf{Mean}& \textbf{Stdev} \\
		\hline\hline
		\Tstrut Exploit & $2$  & $70$ & $3$  & $16.8$ & $24.1$\\
		Binary Download & $2$ & $16$ & $2$ & $4.6$ & $4.6$\\
		C\&C Comm. &  $2$ & $4266$ & $2$ & $18.1$ & $93.2$ \\
		Attack & $2$ & $23003$ & $2$ &  $37.4$ & $694.7$\\
		\hline
	\end{tabular}
	
	\label{Tab:lengthStats}
\end{table}

\begin{table}
	\centering
	\small
	\vspace{1em}
	\caption{Example of Consecutive Attack Alerts}
	\begin{tabular}{| p{0.9\columnwidth}|}
		\hline \Tstrut 
		\texttt{08/17/11-17:16:18.349371  [**] [1:2001581:13] ET SCAN Behavioral Unusual Port 135 traffic, Potential Scan or Infection [**] [Classification: Misc activity] [Priority: 3] {TCP} 147.32.84.130:2888 -> 147.32.82.18:135} \\ \\
		\texttt{08/17/11-17:17:18.436233  [**] [1:2001581:13] ET SCAN Behavioral Unusual Port 135 traffic, Potential Scan or Infection [**] [Classification: Misc activity] [Priority: 3] {TCP} 147.32.84.130:3492 -> 147.32.88.65:135} \\ \hline
	\end{tabular}
	\label{Tab:selfTransSnapshot}
\end{table} 

Before we detail the experiment methodology
for attack prediction using our trained models,
 we discuss an important issue: whether
or not to consider self-transitions in the data.
Self-transitions arise when the underlying IDS generates multiple consecutive alerts for the same state; for example, a host continuously engaging in a port scan attack may generate a stream of scan alerts, which we interpret as a sequence of self-transitions from the  Attack state in our model. Table~\ref{Tab:selfTransSnapshot} shows an example of two consecutive port scan alerts. The rule~\cite{ruleurl} causing the alert is triggered by a threshold of outgoing connections being crossed within a certain time window, in this case if 70 connections are made in 60 seconds. Thus, every 60 seconds, the rule will be triggered again, causing many consecutive alerts for the Attack state if the host is engaging in the activity for an extended period of time, for example several minutes. We observe that consecutive alerts for the same event occur frequently in our dataset; however, this could well be an artefact of using Snort as the IDS. A different IDS may generate only a single alert for the entire duration of a malicious activity. Regardless, we discuss the consequences of self-transitions below.
 
We find that when we build a transition matrix from the original dataset, as shown in Section~\ref{subsec:instantate}, the self-transition probabilities for the C\&C Communication and
Attack states overwhelm the 
probabilities of  transitions leading to other states. 
Thus, when we predict the most likely transition from these states,
we end up always predicting the self-transition; in fact,
the only time we can possibly predict attack is
when the current state is the Attack state. This is not useful, as we need to know when a non-attack state will transition to attack, not whether 
the Attack state will remain in itself. 
Thus we have to consider whether it is possible to drop self-transitions by considering each set of consecutive alerts
as a single alert. Although we are chiefly interested in predicting when a state changes (i.e. when a non-attack state 
transitions to the Attack state) as opposed to its duration (represented by how long it
transitioned to itself), 
it is still important to see whether there is a discernible pattern in the duration or number of each
 state's self-transitions.
 Such a pattern, if it existed, would be
 important, because some non-attack states (say C\&C Communication) may nearly always self-transition for a certain duration before they go to attack; this would allow us to
 to generate a warning that an attack is likely to happen after that certain duration whenever we see the C\&C Communication state. To this end,
 we perform an empirical analysis on our dataset. 

Tables~\ref{Tab:durationStats} and~\ref{Tab:lengthStats} show the statistical properties of the durations (in terms of number of minutes) and number (i.e. number of consecutive alerts) respectively of self-transitions of each state. We find that there is no predictable pattern in either duration or number of transitions. 
The tables show that the standard deviation of both the duration and number of self-transitions is very high compared to the mean for almost all states, and the difference between the minimum and maximum is also generally very large. We do a similar analysis for each individual host, not shown here owing to space constraints, but find that even within individual hosts infected by a single bot, the pattern is unpredictable, with values of both duration and number widely scattered.
Table~\ref{Tab:durationStats} shows that the minimum duration of each state is less than 1 minute, and this is also the modal value for all states.
This, combined with the fact that the actual duration is unpredictable owing to its high variance, leads
us to conclude that whenever we predict an attack following another state, we should conservatively always predict that
it is likely to follow in less than a minute. The actual duration after which it will follow then becomes
irrelevant. 
The downside of this approach is that when an attack does not in reality
follow within a few minutes, but rather after many hours, valuable resources will be wasted in
monitoring the suspected host for a long time, or its communications will be unnecessarily restricted.
However we find this acceptable because long durations of self-transitions are uncommon.
Out of a total of $1233$ occurrences of the Attack state in our dataset, only $27$ times the duration is over $2$ minutes in length. Similarly only $170$ out of $3659$ occurrences of the C\&C Communication state
last longer than $2$ minutes.
Therefore, although self-transitions are part of our original model, we now decide to ignore all self-transitions in the data when training our model, and instead
build our transition matrix
from a dataset where we collapse all self transition sequences into a single state, i.e. a number of 
consecutive alerts for a state will be considered a single alert. In the testing stage,
when we make a prediction from a state, we ignore further alerts for the same state,
until there is a state change, at which point we make another prediction.
Thus, this strategy of ignoring consecutive alerts would work for both kinds of IDS, i.e. those that repeatedly generate a stream of alerts for the same instance of malicious activity (e.g. separate messages exchanged during a single connection with a C\&C server) and those that group all consecutive alerts into a single alert for the activity (e.g. one alert per connection with a C\&C server).

We now obtain the following transition probability matrix, which we call $T'$:

\begin{center}
	$
	T' = 
	\begin{pmatrix}
	0 &	0.0938 &	0.1042 &	0.8021\\
	0.0619 &	0 &	0.9175	& 0.0206\\
	0.0018	& 0.0178 &	0 &	0.9804\\
	0.0154	& 0.0002 &	0.9844 &	0
	\end{pmatrix}
	$
	
\end{center}
As before, the states represented by each row and column are Exploit, Binary Download, C\&C Communication and Attack, respectively.
The values along the diagonal are now $0$ as self-transitions have
been removed. However, it can be verified that this model still satisfies all Markov properties discussed in Section~\ref{subsec:singleMarkov}. All states of the model communicate with each other; hence,
the chain is irreducible. Each state is aperiodic as it can be returned to at time 2,3,4 and so on; thus, the period,
i.e. the GCD (greatest common denominator) of all possible return times, is 1. Finally, the properties of reversibility
and a valid stationary distribution can be verified through calculations as discussed in Section~\ref{subsec:instantate} -- 
we omit the calculations for brevity.

\section{Experimental Methodology and Evaluation Results}
\label{Sec:prediction}
We now investigate the accuracy with which we are able to predict attacks\footnote{The attacks in our evaluation dataset are restricted to sending spam emails and performing outbound scans; while the current results only reflect these two activities, in theory our framework can predict any other attack, as the Attack state could represent one of many different attacks.}. Our prediction methodology is as follows. 

\subsection{Experimental Methodology}
We first divided the dataset temporally into training and testing data - i.e. for each host, we considered the first $85$ minutes of data as training, and the remainder as testing data.
We chose this rather short training interval
as the data duration varies considerably across traces, with some traces lasting only
$2$ to $3$ hours. Therefore, 
using longer training intervals leaves fewer hosts' traces available for testing the model.
We trained and evaluated each model as follows.

\subsubsection{Single-order Markov chain}
 To make predictions using the single-order Markov chain,
we first learn a transition probability matrix from training data, 
similar to the matrix shown in Section~\ref{subsec:instantate}, 
but with different values as it was built on less data.
Then, given a set of states $[S]$, the current state $S_{i}$, and a set of transition probabilities $t_{i,j} \forall j\in S$,
we predict the next state to be the state $S_j$ such that the transition probability 
${
	P(S_i \rightarrow S_j)  =
	\max\limits_{j\in S} t_{i,j}}$.
This means that we predict the Attack state whenever it is the destination of the most likely outgoing transition from the current state. 
We iterate over the testing data, treating it as a real-time, previously unseen stream of events.
After observing each state, we predict whether the next state to follow
was likely to be the Attack state.  Thus, whenever a new state is observed,
we immediately make a prediction for the next state to follow.

\subsubsection{Higher-order Markov chain}
We implement a set of chains from order-2 to order-9, learning a transition probability matrix $T^{(m)}$
for each 
order $m$ which contains probabilities for all four states being preceded by all possible state sequences of length $m$.
In the prediction phase, we iterate over the event stream, using $m$ events to predict the next event. For example, when a new event $E_t$ occurs at time $t$, we look up the row of
the sequence $[E_{t-m},...E_t]$
in the matrix  $T^{(m)}$ and predict the most likely next event $E_{t+1}$.

\subsubsection{Semi-Markov Chain}
We learn eight transition 
probability matrices from the data (not shown for brevity), one for each of the eight time intervals in $[I]$ shown in 
Table~\ref{Tab:intervals}, representing the distribution $F_{ij}(t)$. Each matrix $T_{I_N}$
contains 
the transition probabilities among all states within the time interval $I_N$. 
In the testing phase, we iterate over the testing data, making a prediction as each 
new state occurs, simulating a real-time implementation. At each state change, we not only predict
the next state, but also predict the holding time of the current state.
We calculate the term $H_{S_i}(t_x)$ from Equation~\ref{eq:holdTime} for all $t_x \in [I]$, estimating
a set of probabilities representing the likelihood of the holding time lying within each interval in $[I]$. We
 take the interval with the highest probability as the holding time interval for the current state.
As we make the prediction immediately after a 
state change is observed, the time so far spent in each state when we make the prediction is zero
and is not taken into account.
In our current implementation,
if we predict, for example, that the holding time will lie in interval $3$, it means 
that the next transition is predicted
to occur after $10$ to $20$ minutes(refer to Table~\ref{Tab:intervals}). 
We analyse the error in the holding time prediction for each state in terms of the number of
intervals that the prediction was off: if we predicted the holding time for a state
to lie in interval $3$, but the actual state occurred in interval $5$, the error will be  $Predicted\:Value \thinspace - \thinspace Actual\:Value = -2$, indicating that our prediction
was $2$ intervals early. Positive prediction errors will indicate that we predicted a transition would occur
later than it actually occurred.

\subsection{Predicting Attacks With Single-Order Markov Chain}
\begin{table*}
	\centering
	\small
	\vspace{1em}
	\caption{Accuracy analysis of predicting the Attack and C\&C Communication states.}
	\begin{tabular}{ |p{0.18\textwidth} || p{0.13\textwidth}| p{0.13\textwidth} | p{0.13\textwidth} | p{0.15\textwidth}  | p{0.11\textwidth} | }
		\hline  \Tstrut
		\textbf{State} & \textbf{True Positives} & \textbf{False Positives} & \textbf{True Negatives} & \textbf{False Negatives} & \textbf{Accuracy} \\
		\hline \hline \Tstrut
		Attack & 98.3\% & 1.3\% & 98.7\% & 1.7\% & 98.5\% \\
		C\&C Communication & 99.8\% & 1.7\% & 98.3\% & 0.02\% & 99\% \\
		\hline
	\end{tabular}
	\label{Tab:accuracyCombined}
\end{table*}
\begin{figure*}
	\centering
	\begin{minipage}{.35\textwidth}
		\centering
		\includegraphics[width=\columnwidth]{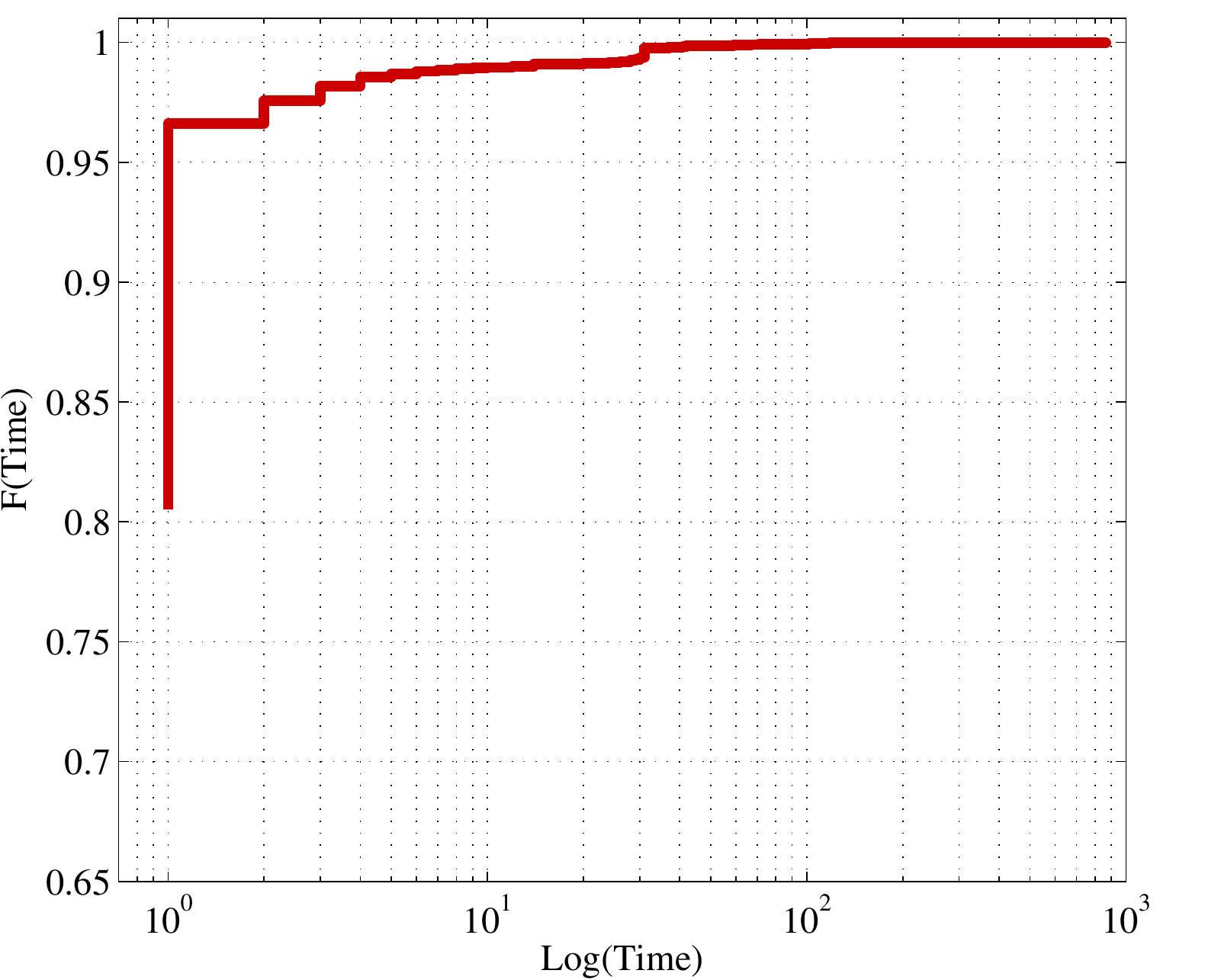}
		\captionof{figure}{CDF of minutes between an attack prediction and the attack itself.}
		\label{fig:attackTimeLag}
	\end{minipage} %
	\begin{minipage}{.35\textwidth}
		\centering
		\includegraphics[width=\columnwidth]{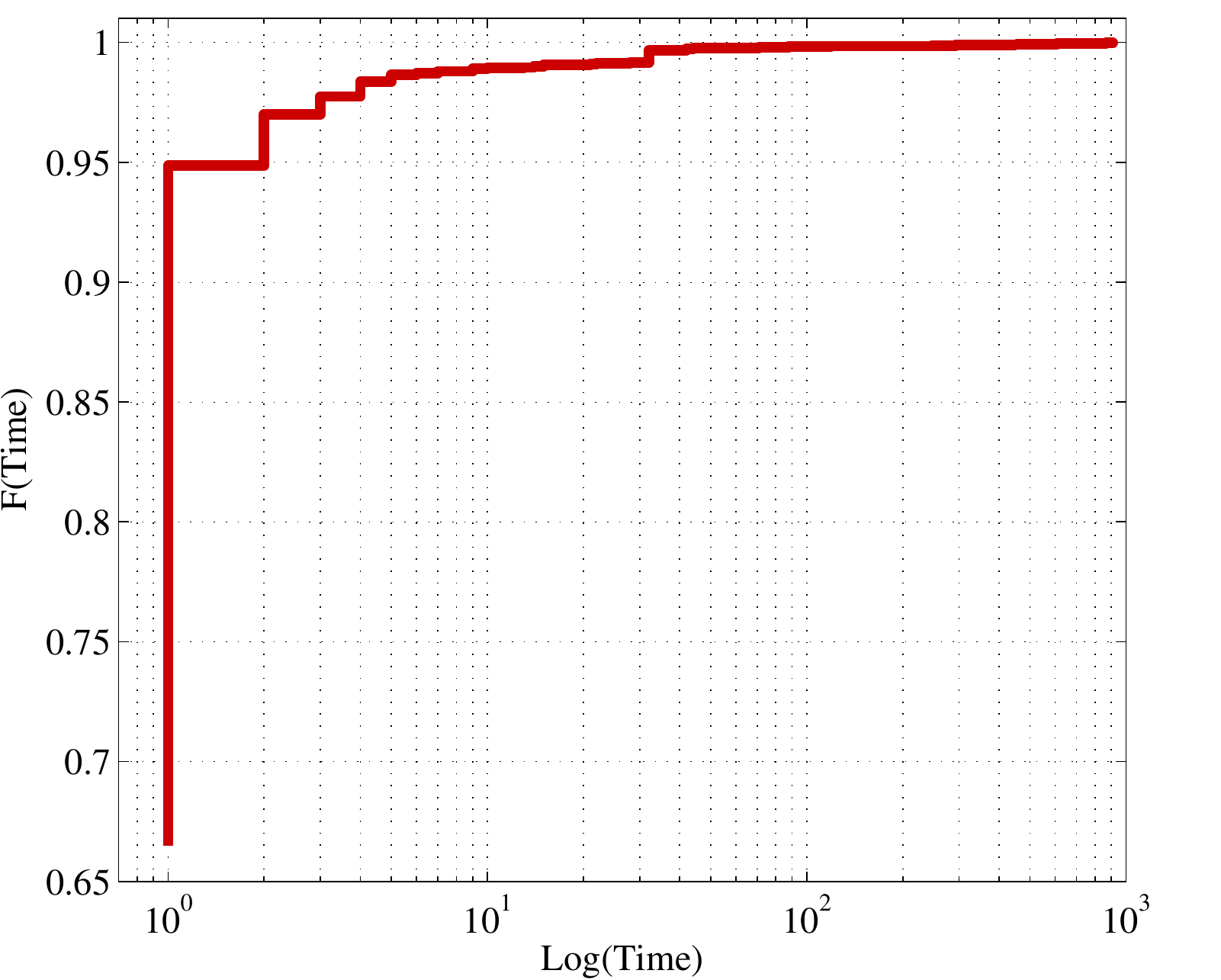}
		\captionof{figure}{CDF of minutes between a C\&C Communication prediction and an attack.}
		\label{fig:cncTimeLag}
	\end{minipage}
\end{figure*}

We now present the results for predicting attacks with the single-order Markov chain.
The accuracy analysis of the experiment is shown in the first row of Table~\ref{Tab:accuracyCombined}. We achieve a very high percentage ($98.3\%$) of true positive predictions with a very low false alarm 
rate of $1.3\%$; i.e. only $1.3\%$ of the time when we made an attack prediction, the attack did not occur
as the immediate next transition. In this experiment, we achieve an overall accuracy of $98.5\%$. 

The earlier that predictions of future attacks can be made, the more useful they are.  Fig.~\ref{fig:attackTimeLag} plots the frequency distribution of the
time that elapsed after each correct attack prediction 
until the Attack state actually occurred as a CDF.
The x-axis is log-scaled to magnify the variation in values smaller than $10$ minutes.
In $81\%$ of cases, the warning time is less than $1$ minute, i.e. after the majority of predictions, less than one minute elapses before an attack is seen. In the remaining $19\%$ of the cases, the warning time ranges from one minute and up to $865$ minutes. 
While in an  automated deployment, a full minute is sufficient to take
 defensive measures against hosts soon to engage in attacks, warning times of only a few
 seconds may be insufficient to apprehend attacks. In such scenarios, our attack prediction is not meeting the desired
objective of having ample early warning to allow for selective defensive measures; 
we now investigate the possibility of increasing
 the warning time by making the prediction even earlier.

\subsection{Predicting C\&C Communication With Single-Order Markov Chain}
We now investigate whether it is possible to generate attack warnings significantly
earlier by predicting states that are very likely to precede the Attack state. 
From
 the transition matrix $T'$ from Section~\ref{sec:selfTransitions}, we can see that
the C\&C Communication State, represented by the third row, is the most likely ($0.98$) to
precede attacks. We hypothesise that 
predicting the C\&C Communication state (instead of attack) would allow for earlier warning of attacks.
In the remainder of this section, we investigate (a) how accurately we can predict the C\&C Communication state, (b) how often the predicted C\&C Communication state is actually preceding the Attack state, and (c) how much of an earlier warning we can get for attacks when 
predicting C\&C Communication instead of predicting the Attack state directly.

As before, we predict
the next state to be C\&C Communication when it is the most likely outgoing transition, achieving $99.8$\% true positives and $1.7$\% false positives (Table~\ref{Tab:accuracyCombined}, row 2). 
Further analysis of the true positives shows that $99\%$ of the time the C\&C Communication state was 
predicted, it was indeed
followed by the Attack state, demonstrating that it is a good indicator of future attacks.
Finally, Fig.~\ref{fig:cncTimeLag} plots the distribution of the time difference between predicting the C\&C Communication state and the occurrence of the 
Attack state. It shows that the proportion of the
cases where we have under a minute of warning reduces from $81\%$ when we were only predicting the Attack state to $66\%$ now that we are able to 
predict the C\&C Communication state. 
While we acknowledge that further improvement is needed, the results do validate
our hypothesis that predicting further back in the state sequence can increase
the warning time before an attack. We hypothesise that for even earlier warning, we can predict the state that most likely leads to the C\&C Communication state.
However, in our current dataset, we only
see $102$ occurrences of the Binary Download state
and $96$ of the Exploit state, compared to nearly $5000$ occurrences of both the attack and C\&C states. 
Clearly this dataset is insufficient 
to  investigate the temporal advantage of predicting these states.

\emph{Comparison with BotHunter:}\\
As no current publicly available tool predicts botnet attacks, we are unable to directly compare our results with any other solutions. However, Figures~\ref{fig:attackTimeLag} and \ref{fig:cncTimeLag} actually represent the temporal advantage offered by our model compared to BotHunter, i.e. the CDF of the time, in minutes, that we are ahead of the time that BotHunter first becomes aware of an attack. BotHunter processes alerts from the Snort IDS (same as our system) and correlates them to identify botnet infections in the network. The earliest time that an attack can become known in a network monitored by BotHunter is when a Snort alert is issued after seeing an attack. Our model on the other hand attempts to predict the attack based on past behaviour. Thus, our system will either be unable to predict attacks and see them for the first time when a Snort alert is generated (same as BotHunter), or predict them earlier than the first Snort alert for them; the time advantage from our system will necessarily be $\ge 0$ and we can never be late in becoming aware of an attack compared to BotHunter. 

\subsection{Predicting with Higher-Order Markov Chains}
\label{Sec:higherOrder}

\begin{figure*}
	\centering
	\begin{minipage}{.35\textwidth}
		\centering
		\includegraphics[width=\columnwidth]{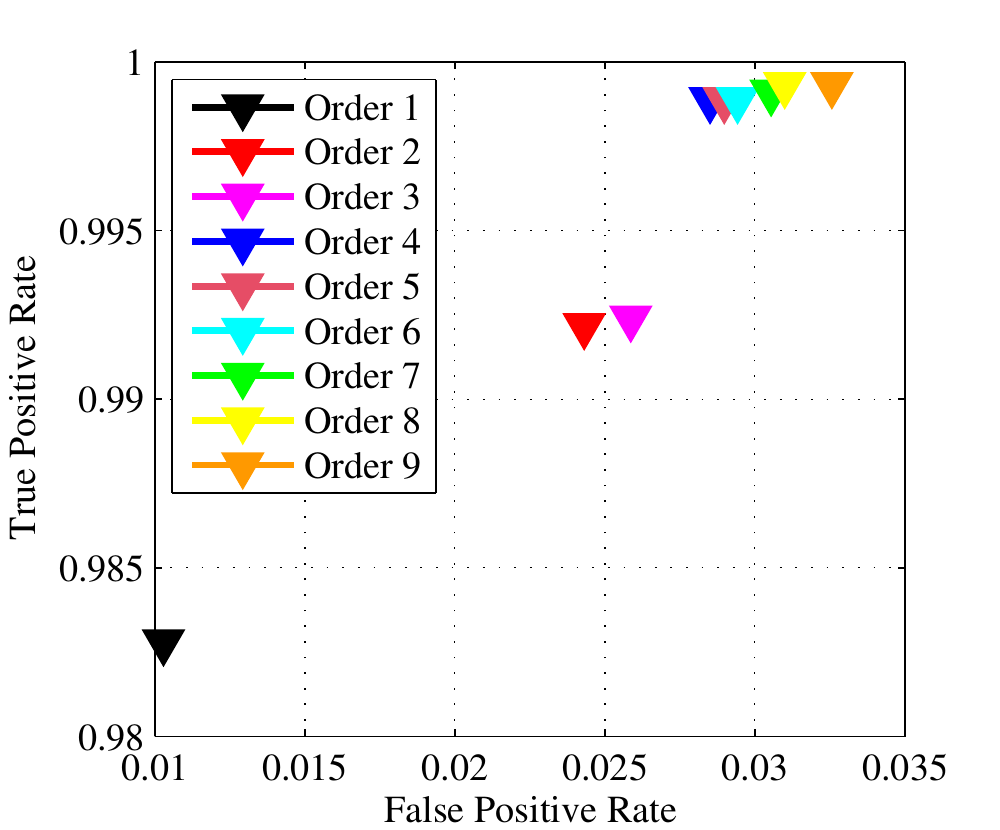}
		\captionof{figure}{TPR and FPR of prediction with higher-order models.}
		\label{fig:higherOrderRoc}
	\end{minipage}%
	\begin{minipage}{.35\textwidth}
		\centering
		\includegraphics[width=\columnwidth]{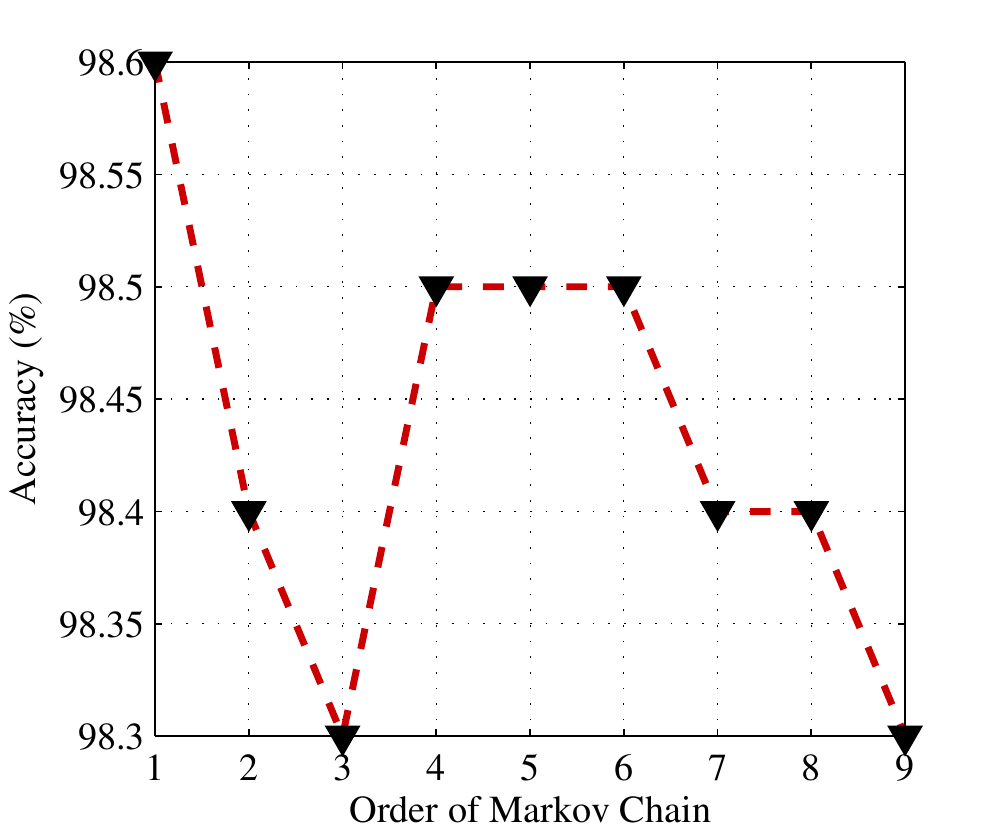}
		\captionof{figure}{Accuracy of prediction with higher-order models.}
		\label{fig:higherOrderAccuracy}
	\end{minipage}%
\end{figure*}

We now extend our single-order Markov chain to higher orders and compare the  difference in accuracy.
Fig.~\ref{fig:higherOrderRoc} shows the true positive rate (TPR) and false positive rate (FPR)
of attack prediction using Markov chains
of order 1 to order 9.  We observe that from order 1 to order 2, the TPR increases
from 98.3\% to 99.2\%, and for each subsequent higher order, the TPR continues to increase, approaching nearly a hundred percent at order 9 (no further improvement in accuracy was
observed for higher orders). However, the improvement in TPR happens at the cost of a 
continual increase in FPR,  and as Fig.~\ref{fig:higherOrderAccuracy} shows, the highest overall accuracy of 98.6\% is achieved by the order-1 Markov chain. Although the decrease in accuracy for higher orders is very slight,
and fluctuates rather than following an exactly linear trend, our key observation is that 
the accuracy of the order-1 chain is not achieved by any higher order. 

 \begin{table*}
	\centering
	\small
	\vspace{1em}
	\caption{Frequency distribution of error in holding time prediction for each state.}
	\begin{tabular}{cc cc cc cc}
		\toprule
		\multicolumn{2}{c}{\textbf{Exploit}} &
		\multicolumn{2}{c}{\textbf{Binary Download}} &
		\multicolumn{2}{c}{\textbf{C\&C Communication}} &
		\multicolumn{2}{c}{\textbf{Attack}} \\
		\midrule
		{Error} & {Frequency} & {Error} & {Frequency} & {Error} & {Frequency} & {Error} & {Frequency}\\
		\midrule
		$-7$ & $0$ & $-7$ & $0$ & $-7$ & $6$ & $-7$ & $10$ \\
		$-6$ & $0$ & $-6$ & $0$ & $-6$ & $1$ & $-6$ & $1$ \\
		$-5$ & $0$ & $-5$ & $0$ & $-5$ & $3$ & $-5$ & $0$ \\
		$-4$ & $0$ & $-4$ & $0$ & $-4$ & $19$ & $-4$ & $1$ \\
		$-3$ & $1$ & $-3$ & $0$ & $-3$ & $13$ & $-3$ & $0$ \\
		$-2$ & $1$ & $-2$ & $0$ & $-2$ & $11$ & $-2$ & $1$ \\
		$-1$ & $3$ & $-1$ & $0$ & $-1$ & $112$ & $-1$ & $45$ \\
		$\:\:\:0$ & $79$ & $\:\:\:0$ & $40$ & $\:\:\:0$ & $4285$ & $\:\:\:0$ & $4423$ \\					
		\hline
	\end{tabular}
	\label{Tab:freqErrors}
\end{table*}

From this investigation, we conclude that the higher-order chain approach does not add a significant benefit for the botnet attack prediction problem in practice. In fact, our results seem to corroborate the conclusion of earlier studies which suggest that a first-order Markov chain has comparable accuracy to higher-order chains in the intrusion detection domain~\cite{schonlau2001computer}. However, because we do observe a clear improvement in TPR that grows with the order of the chain, we argue that some settings can benefit from higher-order chains. For example, in safety-critical networks where attacks are to be strictly prevented, full detection may be desired even at the cost of increased false alarms. It is also important to consider the run-time performance of a high-order chain compared to a single-order chain. As the size of a Markov chain grows as a function of the order of the chain (in general, an $m^{th}$-order Markov chain with $N$ states has $(N-1)N^m$ parameters), the computational complexity of training and predicting with a higher-order chain grows exponentially as the order increases. We discuss the run-time performance of higher-order chains on our dataset in Section~\ref{sec:feasibility}.
%

\subsection{Estimating Attack Occurrence Times using a Semi-Markov Chain}
\label{Sec:semiMarkov}

\begin{figure*}
	\centering
	\begin{minipage}{.35\textwidth}
		\centering
		\includegraphics[width=\columnwidth]{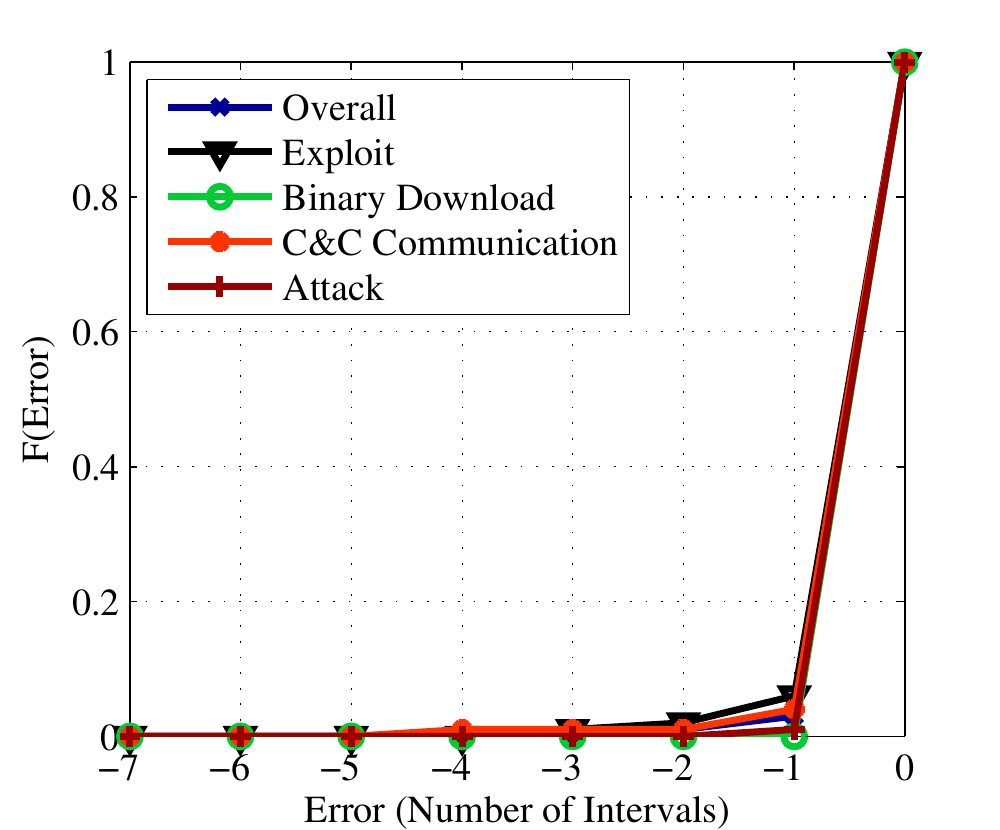}
		\captionof{figure}{CDF of time prediction error for all four states.}
		\label{fig:CDF_All}
	\end{minipage}%
	\begin{minipage}{.35\textwidth}
		\centering
		\includegraphics[width=\columnwidth]{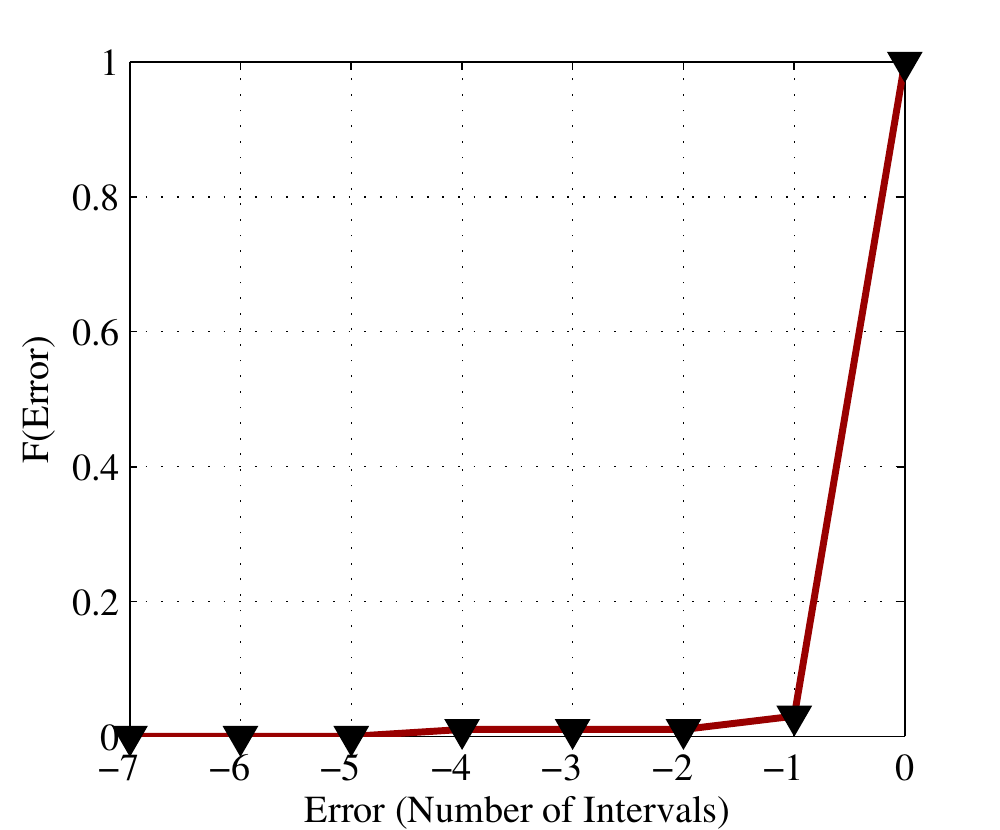}
		\captionof{figure}{CDF of time prediction error for states preceding attacks.}
		\label{fig:CDF_Attack}
	\end{minipage}%
\end{figure*}

We now show the results of predicting the holding time using our semi-Markov chain 
in Fig.~\ref{fig:CDF_All}  and~\ref{fig:CDF_Attack}. 
Each figure
shows a CDF of the prediction error in terms of the number of intervals that our time predictions were off. Fig~\ref{fig:CDF_All} plots the average error in predicting holding times for all states, and
the errors for each state individually.
Table~\ref{Tab:freqErrors} shows the frequency distribution corresponding to Fig~\ref{fig:CDF_All}, for a closer examination
of the holding time prediction errors for each state.
The best possible achievable result
is that a hundred percent of the predictions have zero error, i.e. the next state actually occurred within
the time interval that 
we predicted. 
Our result shows plots that lie close to this theoretical maximum. The
 plot for the Binary Download state in Fig.~\ref{fig:CDF_All}
 shows 100\% correct predictions of holding time 
 and that of  the Attack state shows 99\% correct predictions (as
 the cumulative frequency
 up to the error value $-1$ is $0.01$, or only $1$\%). 
 The proportion of correct
 predictions for holding times of the  
  C\&C Communication
and Exploit states  follow closely, at 96\% and 94\% respectively. On average, 
we make 97\% accurate holding time predictions.
Table~\ref{Tab:freqErrors} shows the number of observations of each state for which holding times
were incorrectly predicted: for all states,
 the error value $-1$ has the maximum frequency -- i.e. most of our incorrect predictions
 were only $1$ interval (i.e. at most $10$ minutes) early compared to the actual occurrence time of the transition. However, even this small prediction error occurred in a very small proportion of instances.

Fig~\ref{fig:CDF_Attack} shows the holding time prediction errors only for those states that precede
the Attack state. This is important because predicting the holding time accurately is most important
when the next transition is to the Attack state, as it tells us how much time we have to contain the
upcoming attacks. The figure shows that the cumulative frequencies of the error values $-7$ and $-6$ are
zero; thus, the worst-case error is $5$ intervals. The cumulative frequency of the error value $-1$ is
0.03, indicating a total error of 3\%, i.e. 
97\% accuracy in predicting the
holding time of the states preceding attack, or in other words, 
predicting the time 
of occurrence of the Attack state.
  Across all our results, we observe that the 
prediction error never has a positive value, which would indicate that transitions
occurred earlier than we predicted them to occur; that is,
 in the small proportion of instances where our prediction is incorrect,
we are predicting transitions \textit{early} rather than late. This implies that 
whenever the next state is predicted
to be an attack, the time prediction may lead to unnecessary preventative measures but
will never lead to a delayed response that allows attacks to occur before the
preventative measures are taken.

\begin{table*}
	\centering
	\small
	\vspace{1em}
	\caption{Analysis of runtime performance of single-order, higher-order and semi-Markov chains.}
	\begin{tabular}{ |p{0.2\textwidth} || p{0.3\textwidth}| p{0.3\textwidth} |}
		\hline  \Tstrut
		\textbf{Model} & \textbf{Training Time} (ms) & \textbf{Per-prediction Time} ($\mu s$) \\
		\hline \hline \Tstrut
		Order-1 Markov Chain & 21 & 0.15 \\
		Order-9 Markov Chain & 3861 & 0.19 \\
		Semi-Markov Chain & 246 & 2.55 \\
		\hline
	\end{tabular}
	\label{Tab:runtime}
\end{table*}

Although the time estimation for our dataset is very accurate, a limitation in our current dataset or IDS limits the usefulness of this time prediction capability. Section~\ref{sec:selfTransitions} showed that most attack and C\&C communication events in the training data occur within the first interval (i.e., within a minute of previous events). This is reflected in the learned $F_{ij}(t)$ distribution, which is used to make holding time predictions. As a consequence, the semi-Markov chain is usually predicting the first interval ($ 0 < t \leq 1$ ) as the time when the next transition will occur, which is insufficient time for manual investigation and in practice would lead to an immediate quarantine response to attack threat. Thus, the advantage of flexibly deciding whether to take a manual or auto-blocking response is lost. However, we argue that this is an artefact of the data  or IDS used in our experiments, where attack events closely follow C\&C communication events. If for example the IDS was generating C\&C communication alerts only once at the beginning of an instance of C\&C communication instead of repeatedly, then attacks would likely be spaced farther apart from the last instance of C\&C communication. Similarly, some bots could deliberately inject delays before generating attacks in response to C\&C instructions, in order to remain stealthy. Thus, with a different dataset or IDS, the time distribution could vary considerably,  and the prediction would not always be the minimum interval, making it important to learn it from data when predicting the times of attacks. Therefore, despite this limitation, we believe that the semi-Markov chain module should still be implemented within a real deployment of our approach. Even if real traffic is similar to our dataset, the semi-Markov chain does not add any significant computational overhead at run-time (as shown in Section~\ref{sec:feasibility}) and does not affect the accuracy of state prediction.



\section{Practical Issues}
\label{sec:feasibility}

\emph{Feasibility:}
We believe our approach is feasible to deploy in practice because it requires no prior knowledge of infection of monitored hosts; any standard IDS can be deployed over live traffic and the Markov model (previously trained on labelled IDS output) can run over the IDS event stream. Because of the simplicity of the approach, it is easy to re-train the model or update it with new states. Re-training on a new dataset can be performed offline and the previously trained model can simply be replaced without any changes in the IDS or prediction models. Updating the model with new states (for example adding more types of attack states) is also simple; the IDS would be updated or replaced (for example, new rules can be added to Snort to allow detecting the new states), and the training and prediction modules would only require minor offline re-programming to allow for a changed number of states. 

\emph{Run-time Performance:}
Markov chains' training time depends on the size of the training data and the order of the chain, as the number of transition probabilities to be learned grows as a function of the order of the chain and number of states. To analyse training time on our dataset, we have trained an order-1 chain, an order-9 chain, and a semi-Markov chain on our complete dataset. Table~\ref{Tab:runtime} shows that the training phase for the order-1 chain and the semi-Markov chain can be completed very quickly (in 21 and 246 milliseconds respectively) because of the simplicity of the 4-state model, while higher-order chains take longer to train (about 3.9 seconds for our order-9 chain). A single prediction with all three model variations is fast enough ( $<3 \mu s$) that they can operate live even in high speed networks. 

\emph{Robustness to Deliberate Evasion:}
An attacker may introduce random delays in compromised hosts' communication or vary the order of the malicious actions, for example by inserting a binary download between the C\&C communication and attack phases of a bot binary. While random delays will not affect the Markov chain's attack prediction, which is time-independent, they will affect the semi-Markov chain's time estimates. Varying the order of the malicious actions will clearly also degrade the prediction accuracy of the model as the live traffic will no longer match the traffic the model was trained on. While designing defences is beyond our current scope, one option is to design a feedback loop into the system, so that live traffic streaming into the system is fed back into the training module after a fixed short time period, for example every hour, when it has been labelled by the IDS, updating the model to reflect the latest malicious behaviour.

\section{Conclusion and Future Work}
\label{sec:concl}
We have presented an approach for predicting the future behaviour of botnet infections,
hypothesising that botnet
attacks can be predicted 
using the malicious behaviour that an infected host will engage in prior to
launching an attack. We have designed a single-order Markov chain, a set of higher-order Markov
chains, and a semi-Markov chain to capture this behavioural sequence
and instantiated the models using real-world botnet traffic. All models showed high prediction
accuracy, with over $98\%$ of attacks predicted accurately and a worst-case
FPR of $1.3\%$ by the single-order Markov chain, and $97$\% accurate predictions
of the times of future attacks by the semi-Markov chain; the latter is a novel capability that
in practice can allow building a time-sensitive threat response
system in which the response to attack warnings can be tailored according
to the amount of time available. We also showed
that the computationally simplest single-order chain achieves higher overall
prediction accuracy compared to higher orders.

We acknowledge some limitations in our current work. 
As acknowledged in Section~\ref{subsec:modelLimitation}, the range of malicious behaviour that we could cover in our evaluation was limited by currently available intrusion detection tools. 
While
our work has served to validate the proposed approach, replicating our experiments with the complete
state model of  Fig.~\ref{fig:stateDiagra} remains for future work.
Secondly, the type of attack currently has no bearing on the prediction; both
spam and port scan are mapped to the same state in the model. With a dataset including more attack types, we would like to map different
attacks
to different states and investigate whether each
attack is preceded by unique behavioural patterns. Finally,
an investigation of adversarial attacks possible against the system and counter-measures is also left as future work.

\balance
\addcontentsline{toc}{section}{Bibliography}
\bibliographystyle{abbrv} 
\bibliography{archiv_feb21}

\begin{IEEEbiography}[{\includegraphics[width=1in,height=1.25in,clip,keepaspectratio]{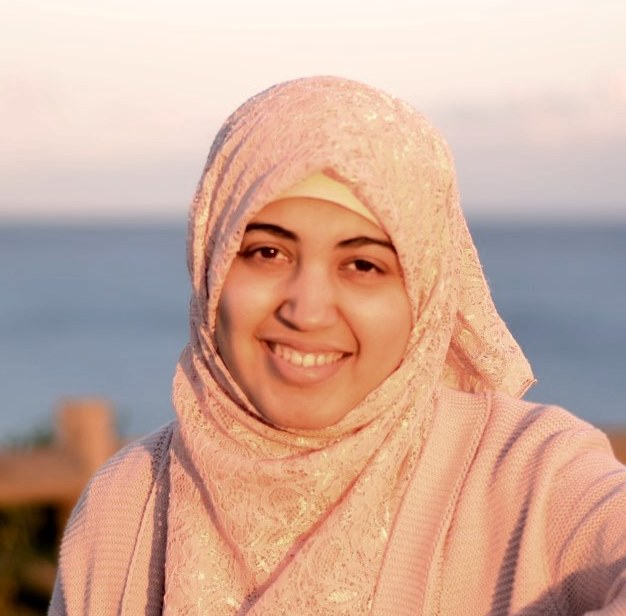}}]{Zainab Abaid} received her Ph.D. from The University of New South Wales, Australia and is currently engaged in a research role with CSIRO Data61, Australia. Her research interests include malware detection and adversarial machine learning. 
\end{IEEEbiography} 
\vspace{-7mm}
\begin{IEEEbiography}[{\includegraphics[width=1in,height =1.25in,clip,keepaspectratio]{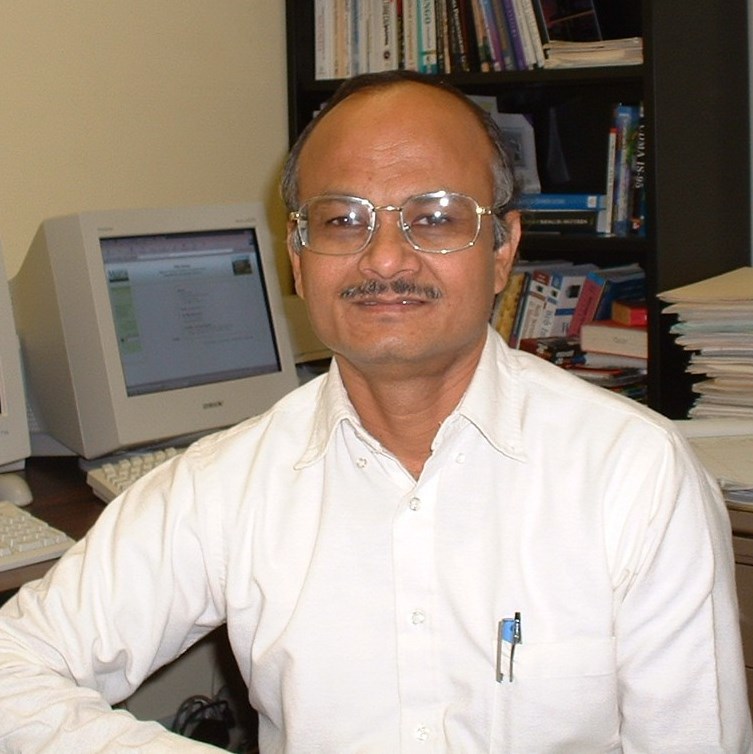}}]{Dilip Sarkar} (SM'96) received his
	Ph.D. from the University of
	Central Florida. He is an Associate
	Professor of Computer Science at the
	University of Miami, Coral Gables.
	His research interests include VBR
	video traffic modeling, middleware and
	Web computing, and parallel and distributed
	processing. He is a senior member of the IEEE, a member of
	IEEE Computer Society and the Association for Computing
	Machinery
\end{IEEEbiography}

\vspace{-7mm}
\begin{IEEEbiography}[{\includegraphics[width=1in,height =1.25in,clip,keepaspectratio]{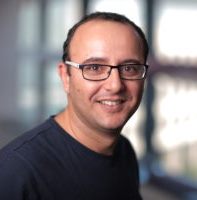}}]{Mohamed Ali Kaafar} received his Ph.D. from INRIA Sophia Antipolis and
	is a Full Professor at the Faculty of Science and Engineering at Macquarie University, the scientific Director of the Optus-Macquarie University Cyber Security Hub, and the group leader of the Information Security and Privacy research group at CSIRO Data61. He is the associate editor of ACM Transactions on Modeling and Performance Evaluation of Computing Systems (ACM Tompecs) and his research interests include Privacy Preserving Technologies, Networks Security, malware detection and Applied Cryptography.
\end{IEEEbiography}
\vspace{-7mm}
\begin{IEEEbiography}[{\includegraphics[width=1in,height =1.25in,clip,keepaspectratio]{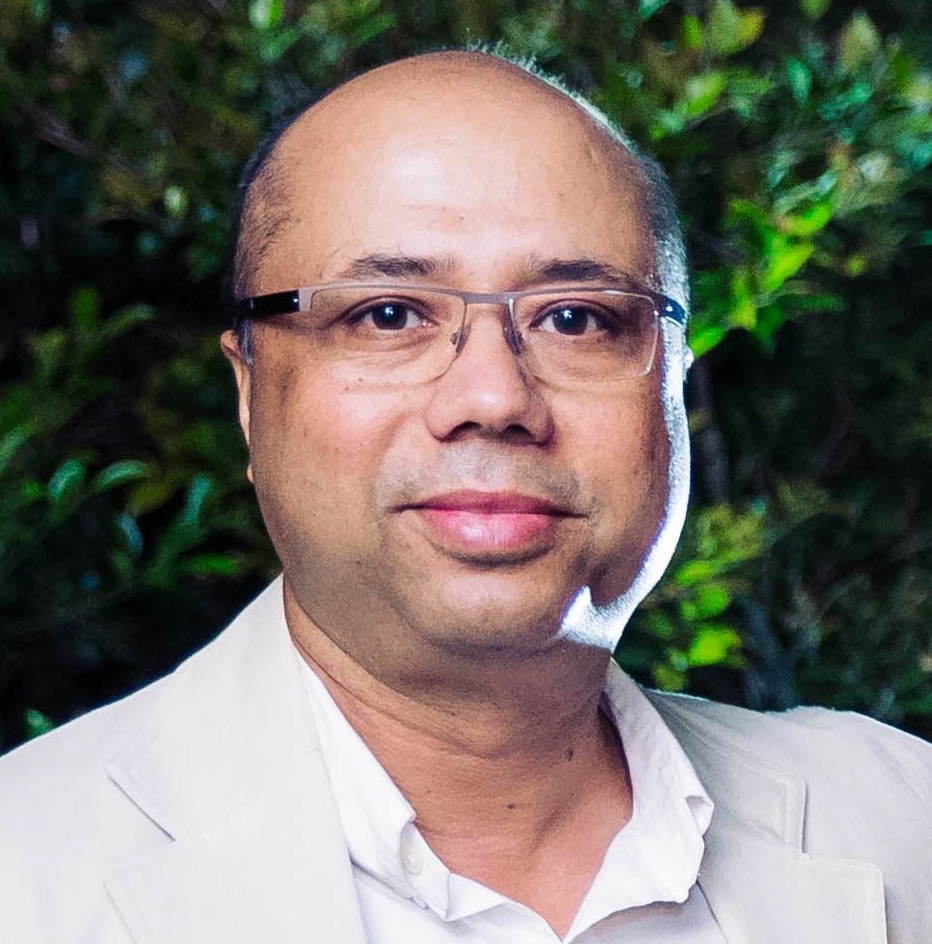}}]{Sanjay Jha} 
	received his Ph.D. from the
	University of Technology, Sydney, Australia. He
	is head of the Networks Group at
	the School of Computer Science and Engineering
	at the University of New South Wales.
	His research activities include a wide range of topics in networking
	including wireless sensor networks, ad hoc/community wireless
	networks, resilience/quality of service (QoS) in IP networks, and active/
	programmable networks. He is a senior member of the
	IEEE and the IEEE Computer Society.
\end{IEEEbiography}

\end{document}